\documentclass[12pt,fleqn,a4paper]{article}
\usepackage{amssymb}
\usepackage{amsmath}
\usepackage[dvips]{graphicx}
\begin{document}
\begin{flushright}
KEK-TH-808 \\
{\tt hep-ph/0203110 }\\
\end{flushright}
\vspace*{1.5cm}
\begin{center}
    {\baselineskip 25pt
    \Large{\bf 
    Detailed calculation of lepton flavor violating muon-electron 
    conversion rate for various nuclei
    
    }
    }

    \vspace{1.2cm}
    \def\thefootnote{\fnsymbol{footnote}}
    {\large Ryuichiro Kitano,$^{\!\!ab,}$\footnote
    {email: {\tt ryuichiro.kitano@kek.jp}}
    Masafumi Koike,$^{\!\!a,}$\footnote
    {email: {\tt mkoike@post.kek.jp}} and
    Yasuhiro Okada$^{ab,}$\footnote
    {email: {\tt yasuhiro.okada@kek.jp}}
    }
    \vspace{.5cm}
    
    {\small {\it $^a$Theory Group, KEK, Oho 1-1, Tsukuba, 
    Ibaraki 305-0801, Japan \\
    \vspace*{2mm}
    $^b$Department of Particle and Nuclear Physics,
    The Graduate University for Advanced Studies,\\
    Oho 1-1, Tsukuba, Ibaraki 305-0801, Japan}}
    
    \vspace{.5cm}
    \today
    
    \vspace{1.5cm}
    {\bf Abstract}
\end{center}

\bigskip

The coherent $\mu$-$e$ conversion rates in various nuclei are
calculated for general lepton flavor violating interactions.
We solve the Dirac equations numerically for the initial state muon
and the final state electron in the Coulomb force, and perform the
overlap integrals between the wave functions and the nucleon
densities.
The results indicate that the conversion branching ratio increases for
a light nucleus up to the atomic number $Z \sim 30$, is largest for $Z
= 30$ -- $60$, and becomes smaller for a heavy nucleus with $Z \gtrsim
60$.
We also discuss the uncertainty from the input proton and neutron
densities.
The atomic number dependence of the conversion ratio calculated here
is useful to distinguish theoretical models with lepton flavor
violation.


\newpage
\def\thefootnote{\arabic{footnote}}
\setcounter{footnote}{0}
\baselineskip 20pt



\section{Introduction}

The observation of lepton flavor violation (LFV) is one of the most
interesting signals beyond the Standard Model (SM).
The charged-lepton LFV processes such as the $\mu \to e \gamma$ decay
and the $\mu$-$e$ conversion in muonic atoms can occur in many 
promising
candidates beyond the SM, although the simple seesaw neutrino model
does not induce experimentally observable rate for the $\mu \to e
\gamma$ process.
For example, sleptons in the supersymmetric (SUSY) extension of the SM
and bulk neutrinos in the higher dimensional models generate LFV
processes through one-loop diagrams \cite{Kuno:1999jp,Faraggi:1999bm}. 
In the R-parity violating SUSY models, additional LFV interactions
exist at the tree level \cite{models-R-parity}.
The branching ratios of the LFV processes have been calculated in many
models in the literature, especially for supersymmetric grand unified
theories (SUSY-GUTs) \cite{Barbieri:1995tw,models-GUTs} and a SUSY
model with right-handed neutrinos \cite{models-Right-Handed-nu}.
It was shown that $\mu \to e \gamma$ and $\mu$-$e$ conversion
branching ratios can be close to the experimental bounds in these
models.

There are on-going and planned experiments for the $\mu \to e \gamma$
and $\mu$-$e$ conversion searches.
For the $\mu \to e \gamma$ branching ratio, the present upper bound is
$1.2 \times 10^{-11}$ from the MEGA collaboration
\cite{Brooks:1999pu}.  A new experiment is under construction at PSI
aiming for a sensitivity of $10^{-14}$ \cite{PSI}.
For the $\mu$-$e$ conversion, an upper bound for the conversion
branching ratio is $6.1 \times 10^{-13}$ \cite{SINDRUM-II} reported by
the SINDRUM II experiment at PSI. Now SINDRUM II is running with gold
(Au) targets.
The MECO experiment at BNL \cite{MECO} are planned in order to search
for $\mu {\rm Al} \to e {\rm Al}$ with a sensitivity below $10^{-16}$.
In future, further improvements by one or two orders of magnitude are
considered for both $\mu \to e \gamma$ and $\mu$-$e$ conversion
processes in the PRISM project \cite{PRISM} at the new 50 GeV proton
synchrotron constructed as a part of the JAERI/KEK joint project.

In order to compare the sensitivity to the LFV interaction in various
nuclei, precise calculation of the $\mu$-$e$ conversion rate is
necessary.
There have been several calculations of the conversion rate.
Weinberg and Feinberg calculated in the case that the conversion
occurs through the photonic interactions ($\mu$-$e$-$\gamma$ vertex)
\cite{WF}.  In the calculation, they used several approximations in
which the muon wave function was taken to be constant in a nucleus and
the outgoing electron was treated as a plane wave.  The plane wave
treatment of the electron is a good approximation only for light
nuclei because the effect of Coulomb distortion on the electron wave
function is large for heavy targets.
The non-photonic interaction case was studied by Marciano and 
Sanda \cite{Marciano:cj}.
Shanker improved the calculation by solving the Dirac equations for
the muon and electron wave functions in the electric potential of a
nucleus \cite{Shanker:1979ap}.  The calculation was carried out for
all the interactions including the photonic and four-fermion operators
in the effective Lagrangian, but the treatment of the photonic dipole
operator was incomplete because he used an approximation that
off-shell photon exchange was replaced by the four-fermion
interaction.
Recently, Czarnecki {\it et al.} presented calculation in which the
off-shell photon is properly treated as an electric field in a nucleus
and listed the values of the conversion rate for aluminum (Al),
titanium (Ti), and lead (Pb) targets \cite{Czarnecki:1998iz}.
The transition rate to the ground state of a nucleus as well as
excited states are calculated in
Refs.\cite{Z-dependence,Kosmas:2001ij}.

In this paper, we evaluated the $\mu$-$e$ conversion rates for nuclei
of a wide range of atomic numbers by the method of Czarnecki {\it et
al.} We took into account all the operators for the $\mu$-$e$
transition.
For any types of operators, the results of our calculation indicate a
tendency that the conversion branching ratio is larger for the nuclei
with moderate atomic numbers than that for light or heavy nuclei.
Although the tendency is the same, there are significant differences
in $Z$ dependences of the conversion rate for various LFV couplings.
The experiments in various nuclei are therefore useful for model
discrimination because each theoretical model predicts different $Z$
dependences.
The conversion rate depends on the input proton and neutron densities
for each nucleus.  Although the proton density is well measured by
electron scattering, there is uncertainty in the determination of
neutron densities.
We estimate the uncertainty from these input parameters.  Based on
neutron density distribution determined from proton scattering
experiments performed in 1970's and pionic atom experiments, the
conversion rate changes by 20\% -- 30\% for heavy nuclei.  The
ambiguity is shown to be significantly reduced for lead (Pb) by the
improved determination of the neutron density from a new proton
scattering experiment.

This paper is organized as follows.
In section \ref{section:formula}, we present a formula of the
conversion rate with the most general effective Lagrangian for LFV.
The results of our calculation and the estimation of the uncertainty
are shown in section \ref{section:results}.
In section \ref{section:summary}, we summarize this paper.
The model parameters of the nucleon density functions in nuclei
and the muon capture rate in nuclei are listed in Appendices 
\ref{appendix:nucleon-densities} and \ref{appendix:capture-rate},
respectively.


\section{Formula of $\mu$-$e$ conversion rate}
\label{section:formula}

In this section, we present a method of the conversion rate
calculation.
We solve the Dirac equations for the muon and electron in the initial
and final states, respectively, and obtain transition amplitudes by
integrating the overlap of both wave functions.

We start with the most general LFV interaction Lagrangian which
contributes to the $\mu$-$e$ transition in nuclei \cite{Kuno:1999jp}:
\begin{eqnarray}
    {\cal L}_{\rm int} &=&
    - \frac{4 G_{\rm F}}{\sqrt{2}}
    \left(
    m_\mu A_R \bar{\mu} \sigma^{\mu \nu} P_L e F_{\mu \nu}
    + m_\mu A_L \bar{\mu} \sigma^{\mu \nu} P_R e F_{\mu \nu}
    + {\rm h.c.}
    \right) \nonumber \\
    &&
    - \frac{G_{\rm F}}{\sqrt{2}} \sum_{q = u,d,s}
    \left[ {\rule[-3mm]{0mm}{10mm}\ } \right.
    \left(
    g_{LS(q)} \bar{e} P_R \mu + g_{RS(q)} \bar{e} P_L \mu
    \right) \bar{q} q  \nonumber \\
    &&   \hspace*{2.4cm}
    +
    \left(
    g_{LP(q)} \bar{e} P_R \mu + g_{RP(q)} \bar{e} P_L \mu
    \right) \bar{q} \gamma_5 q \nonumber \\
    &&   \hspace*{2.4cm}
    +
    \left(
    g_{LV(q)} \bar{e} \gamma^{\mu} P_L \mu
    + g_{RV(q)} \bar{e} \gamma^{\mu} P_R \mu
    \right) \bar{q} \gamma_{\mu} q \nonumber \\
    &&   \hspace*{2.4cm}
    +
    \left(
    g_{LA(q)} \bar{e} \gamma^{\mu} P_L \mu
    + g_{RA(q)} \bar{e} \gamma^{\mu} P_R \mu
    \right) \bar{q} \gamma_{\mu} \gamma_5 q \nonumber \\
    &&   \hspace*{2.4cm}
    + \ \frac{1}{2}
    \left(
    g_{LT(q)} \bar{e} \sigma^{\mu \nu} P_R \mu
    + g_{RT(q)} \bar{e} \sigma^{\mu \nu} P_L \mu
    \right) \bar{q} \sigma_{\mu \nu} q
	+ {\rm h.c.}
    \left. {\rule[-3mm]{0mm}{10mm}\ } \right]\ ,
    \label{1}
\end{eqnarray}
where $G_{\rm F}$ and $m_\mu$
are the Fermi constant and the muon mass, respectively, and
$A_{L,R}$ and $g$'s are all dimensionless coupling constants
for the corresponding operators.
Our conventions are 
$F_{\mu \nu} = \partial_\mu A_\nu - \partial_\nu A_\mu$, %
$\sigma^{\mu \nu} = (i / 2) [ \gamma^\mu, \gamma^\nu ]$, %
$P_{L} = (1 - \gamma_{5}) / 2$, %
$P_{R} = (1 + \gamma_{5}) / 2$, %
and the covariant derivative is defined as
$D_\mu = \partial_\mu - i Q e A_\mu$, %
where $Q e$ ($e>0$) is the electric charge
($Q=-1$ for the electron and the muon).
The size of each coupling constant depends
on the interaction of the new physics 
in which the lepton flavor conservation is violated.
There are two types of amplitudes for photonic transition
($\mu$-$e$-$\gamma$), namely the monopole and dipole $\mu$-$e$
transitions.  In the above expression, the monopole transition is
converted to the vector-vector interaction assuming that the momentum
dependences of the form factors are negligible.

The initial state in the $\mu$-$e$ conversion process is the $1s$
state of the muonic atom, and the final electron state is the
eigenstate with an energy of $m_\mu - \epsilon_b$, where $\epsilon_b$
is the binding energy of the $1s$ muonic atom.
Both wave functions in
the initial and final states
can be determined by solving 
the Dirac equations in the electric field of the nucleus.
The Dirac equation in the central force system
is given by \cite{ROSE}
\begin{eqnarray}
    W \psi = \left[
    -i \gamma_5 \sigma_r \left(
    \frac{\partial}{\partial r}
    +\frac{1}{r}
    -\frac{\beta}{r} K
    \right) + V(r) + m_i \beta
    \right] \psi,
    \label{2}
\end{eqnarray}
\begin{eqnarray}
    \gamma_5 = \left(
    \begin{array}{cc}
	0 & 1 \\
	1 & 0\\
    \end{array}
    \right)\ ,\ \ 
    \beta = \left(
    \begin{array}{cc}
	1 & 0\\
	0 &  -1\\
    \end{array}
    \right)\ ,\ \ 
    \sigma_r = 
    \left(
    \begin{array}{cc}
	\boldsymbol{\sigma} \cdot \boldsymbol{r}& 0\\
	0& \boldsymbol{\sigma} \cdot \boldsymbol{r} \\
    \end{array}
    \right),\nonumber
\end{eqnarray}
\begin{eqnarray}
    K = \left(
    \begin{array}{cc}
	\boldsymbol{\sigma} \cdot \boldsymbol{l} + 1&0 \\
	0& -( \boldsymbol{ \sigma} \cdot \boldsymbol{l} + 1) \\
    \end{array}
    \right)
    \ ,
\end{eqnarray}
where $W$ and $V(r)$ are the energy and potential, respectively,
$m_{i}$ is the reduced mass of the electron or the muon,
$\boldsymbol{\sigma}$ are the Pauli matrices, and the orbital angular
momentum $\boldsymbol{l}$ is defined by $\boldsymbol{l} \equiv - i
\boldsymbol{r} \times \nabla$.
Since the operator $K$ and the $z$-component of the total angular
momentum $j_{z}$ commute with the Hamiltonian, two eigenvalues of
these operators, $-\kappa$ and $\mu$, represent the quantum numbers of
the wave functions describing this system.
We denote the wave function as follows:
\begin{eqnarray}
    \psi = \psi^\mu_\kappa = \left(
    \begin{array}{c}
	g(r) \chi^\mu_\kappa (\theta, \phi)\\
	i f(r) \chi^\mu_{- \kappa} (\theta, \phi)\\
    \end{array}
    \right),
    \label{3}
\end{eqnarray}
where $\chi^\mu_\kappa$ is the normalized eigenfunction of 
$(\boldsymbol{\sigma} \cdot \boldsymbol{l} + 1 )$ and $j_z$
such as
\begin{eqnarray}
    (\boldsymbol{\sigma} \cdot \boldsymbol{l} + 1 ) \chi^\mu_\kappa 
    = - \kappa \chi^\mu_\kappa\ , \ \ 
    j_z \chi^\mu_\kappa = \mu \chi^\mu_\kappa\ ,\ \ 
    \int_{-1}^{1} d\cos \theta \int_0^{2 \pi} d \phi \ 
    \chi^{\mu *}_\kappa \chi^{\mu^\prime}_{\kappa^\prime}
    = \delta_{\mu \mu^\prime} \delta_{\kappa \kappa^\prime}\ .
\end{eqnarray}
The total angular momentum $j$ is related to $\kappa$ as 
$\kappa = \pm ( j + 1/2 )$.
With the notation of $u_1(r) = r g(r)$ and $u_2(r) = r f(r)$, the
Dirac equation for the radial function is given by
\begin{eqnarray}
    \frac{d}{dr} \left(
    \begin{array}{c}
	u_1\\
	u_2\\
    \end{array}
    \right) = \left(
    \begin{array}{cc} 
	- {\kappa}/{r}  & W - V + m_i \\
	-(W - V - m_i)    &   {\kappa}/{r} \\
    \end{array}
    \right)
    \left(
    \begin{array}{c}
	u_1 \\
	u_2 \\
    \end{array}
    \right)\ .
    \label{5}
\end{eqnarray}
The initial muon state corresponds to the quantum numbers of
$\mu = \pm 1/2$ and $\kappa = -1$
with a normalization of 
\begin{equation}
    \int d^3 x \psi^{(\mu)*}_{1s} ({\bf x})
    \psi^{(\mu)}_{1s} ({\bf x}) = 1 \ .
    \label{7}
\end{equation}
The final electron state is one of the states in the continuum
spectrum.  Our normalization convention is taken as
\begin{eqnarray}
    \int d^3 x
    \psi^{\mu (e) *}_{\kappa, W} ({\bf x})
    \psi^{\mu^\prime (e)}_{\kappa^\prime, W^\prime} ({\bf x})
    = 
    2 \pi
    \delta_{\mu \mu^\prime} \delta_{\kappa \kappa^\prime}
    \delta (W-W^\prime)\ .
    \label{8}
\end{eqnarray}       

The conversion rate $\omega_{\rm conv}$ in this normalization is
simply written by the square of the $\mu$-$e$ conversion amplitude
$M$, taking the spin average of the initial muon and summing over the
final states of the electron.
From the effective Lagrangian (\ref{1}), $M$ is obtained as follows:
\begin{eqnarray}
    M &=& 
    \frac{4 G_{\rm F}}{\sqrt{2}} \int d^3 x 
    \left(
    m_\mu A_R^* \bar{\psi}^{\mu (e)}_{\kappa, W} 
    \sigma^{\alpha \beta} P_R \psi^{(\mu)}_{1s}
    + m_\mu A_L^* \bar{\psi}^{\mu (e)}_{\kappa, W} 
    \sigma^{\alpha \beta} P_L \psi^{(\mu)}_{1s}
    \right)
    \langle N^\prime | F_{\alpha \beta} | N \rangle \nonumber \\
    &&
    + \frac{G_{\rm F}}{\sqrt{2}} \sum_{q = u,d,s} \int d^3 x 
    \left[ {\rule[-3mm]{0mm}{10mm}\ } \right.
    \left(
    g_{LS(q)} \bar{\psi}^{\mu (e)}_{\kappa, W} P_R \psi^{(\mu)}_{1s} 
    + g_{RS(q)} \bar{\psi}^{\mu (e)}_{\kappa, W} P_L \psi^{(\mu)}_{1s}
    \right) \langle N^\prime | \bar{q} q | N \rangle \nonumber \\
    &&   \hspace*{1.5cm}
    +
    \left(
    g_{LP(q)} \bar{\psi}^{\mu (e)}_{\kappa, W} P_R \psi^{(\mu)}_{1s} 
    + g_{RP(q)} \bar{\psi}^{\mu (e)}_{\kappa, W} P_L \psi^{(\mu)}_{1s}
    \right) \langle N^\prime | \bar{q} \gamma_5  q | N \rangle
    \nonumber \\
    &&   \hspace*{1.5cm}
    +
    \left(
    g_{LV(q)} \bar{\psi}^{\mu (e)}_{\kappa, W} \gamma^{\alpha} P_L 
    \psi^{(\mu)}_{1s}
    + g_{RV(q)} \bar{\psi}^{\mu (e)}_{\kappa, W} \gamma^{\alpha} P_R 
    \psi^{(\mu)}_{1s}
    \right) \langle N^\prime | \bar{q} \gamma_{\alpha} q 
    | N \rangle \nonumber \\
    &&   \hspace*{1.5cm}
    +
    \left(
    g_{LA(q)} \bar{\psi}^{\mu (e)}_{\kappa, W} \gamma^{\alpha} P_L 
    \psi^{(\mu)}_{1s}
    + g_{RA(q)} \bar{\psi}^{\mu (e)}_{\kappa, W} \gamma^{\alpha} P_R 
    \psi^{(\mu)}_{1s}
    \right) \langle N^\prime | 
    \bar{q} \gamma_{\alpha} \gamma_5 q | N \rangle \nonumber \\
    &&   \hspace*{1.5cm}
    + \ \frac{1}{2}
    \left(
    g_{LT(q)} \bar{\psi}^{\mu (e)}_{\kappa, W} 
    \sigma^{\alpha \beta} P_R 
    \psi^{(\mu)}_{1s}
    + g_{RT(q)} \bar{\psi}^{\mu (e)}_{\kappa, W} 
    \sigma^{\alpha \beta} P_L 
    \psi^{(\mu)}_{1s}
    \right) \langle N^\prime | \bar{q} \sigma_{\alpha \beta} q 
    | N \rangle
    \left. {\rule[-3mm]{0mm}{10mm}\ } \right]\ ,\nonumber \\
\end{eqnarray}
where $\langle N^\prime |$ and $| N \rangle $
are the final and initial states of the nucleus, respectively.

Hereafter, we concentrate on the coherent conversion processes
in which the final state of the nucleus is the same as the initial 
one.
The fraction of the coherent process is generally larger than
incoherent one approximately by a factor of the mass number of the
target nuclei.
In this case,
the matrix elements of
$\langle N | \bar{q} \gamma_5 q | N \rangle $,
$\langle N | \bar{q} \gamma_\alpha \gamma_5 q | N \rangle $,
and
$\langle N | \bar{q} \sigma_{\alpha \beta} q | N \rangle $
vanish identically,
and 
$\langle N | \bar{q} q | N \rangle $ and
$\langle N | \bar{q} \gamma_\alpha q | N \rangle $
can be expressed by
the proton and neutron densities ($\rho^{(p)}$ and $\rho^{(n)}$) 
in nuclei as follows:
\begin{eqnarray}
	\langle N | \bar{q} q | N \rangle
	=
	Z G_{S}^{(q, p)} \rho^{(p)} +
	(A - Z) G_{S}^{(q, n)} \rho^{(n)}\ ,
\end{eqnarray}
\begin{eqnarray}
	\langle N | \bar{q} \gamma^0 q | N \rangle
	= \left \{
	\begin{array}{ll}
		2 Z \rho^{(p)} + (A - Z) \rho^{(n)}
		\qquad & {\mbox{for}} \quad q = u, \\
		Z \rho^{(p)} + 2 (A - Z) \rho^{(n)}
		\qquad & {\mbox{for}} \quad q = d, \\
		0
		\qquad & {\mbox{for}} \quad q = s, \\
	\end{array}
	\right.
\end{eqnarray}
\begin{eqnarray}
	\langle N | \bar{q} \gamma^i q | N \rangle = 0
	\qquad (i=1,2,3)\ .
\end{eqnarray}
Here we introduce coefficients $G_{S}^{(q, p)}$ and $G_{S}^{(q, n)}$
for scalar operators.  These are evaluated to be $G_{S}^{(u, p)} =
G_{S}^{(d, n)} = 5.1$, $G_{S}^{(d, p)} = G_{S}^{(u, n)} = 4.3$, and
$G_{S}^{(s, p)} = G_{S}^{(s, n)} = 2.5$ in Kosmas {\it et al.}
\cite{Kosmas:2001mv}.
We assume that the proton and neutron densities are
spherically symmetric and normalized as follows:
\begin{eqnarray}
    \int_0^\infty d r 4 \pi r^2 \rho^{(p,n)} (r) = 1\ .
\end{eqnarray}

The final formula of the conversion rate can be written as follows:
\begin{eqnarray}
	\omega_{\rm conv} &=&
	2 G_{\rm F}^2
	\left|
	A_R^{\ast}  D
	+ \tilde{g}_{LS}^{(p)} S^{(p)}
	+ \tilde{g}_{LS}^{(n)} S^{(n)}
	+ \tilde{g}_{LV}^{(p)} V^{(p)}
	+ \tilde{g}_{LV}^{(n)} V^{(n)}
	\right|^2
	\nonumber \\
	& + &
	2 G_{\rm F}^2
	\left|
	A_L^{\ast}  D
	+ \tilde{g}_{RS}^{(p)} S^{(p)}
	+ \tilde{g}_{RS}^{(n)} S^{(n)}
	+ \tilde{g}_{RV}^{(p)} V^{(p)}
	+ \tilde{g}_{RV}^{(n)} V^{(n)}
	\right|^2.
	\label{eq:omega-final}
\end{eqnarray}
$A_{L}$ and $A_{R}$ are given in Eq.(\ref{1}).  The coupling constants
$\tilde{g}$'s in Eq.(\ref{eq:omega-final}) are defined as
\begin{eqnarray}
	\tilde{g}_{LS, RS}^{(p)}
	& = &
	\sum_q G_{S}^{(q, p)} \  g_{LS, RS(q)}  \ ,
	\\
	\tilde{g}_{LS, RS}^{(n)}
	& = &
	\sum_q G_{S}^{(q, n)} \  g_{LS, RS(q)}  \ ,
	\\
	\tilde{g}_{LV, RV}^{(p)}
	& = &
	2 g_{LV, RV(u)} + g_{LV, RV(d)}\ ,
	\\
	\tilde{g}_{LV, RV}^{(n)}
	& = &
	g_{LV,RV(u)} + 2 g_{LV, R(d)}\ .
\end{eqnarray}
We also introduced the following overlap integrals:
\begin{eqnarray}
	D = 
	\frac{4}{\sqrt{2}} m_\mu 
	\int_0^\infty dr r^2 [ -E(r) ] ( g_e^- f_\mu^- + f_e^- g_\mu^- )
	\ ,
	\label{22}
\end{eqnarray}
\begin{eqnarray}
	S^{(p)} =
	\frac{1}{2 \sqrt{2}} 
	\int_0^\infty dr r^2 
	Z \rho^{(p)} ( g_e^- g_\mu^- - f_e^- f_\mu^- )
	\ ,
	\label{23}
\end{eqnarray}
\begin{eqnarray}
	S^{(n)} =
	\frac{1}{2 \sqrt{2}} 
	\int_0^\infty dr r^2 
	(A-Z) \rho^{(n)} ( g_e^- g_\mu^- - f_e^- f_\mu^- )
	\ ,
	\label{24}
\end{eqnarray}
\begin{eqnarray}
	V^{(p)} =
	\frac{1}{2 \sqrt{2}} 
	\int_0^\infty dr r^2 
	Z \rho^{(p)} ( g_e^- g_\mu^- + f_e^- f_\mu^- )
	\ ,
	\label{25}
\end{eqnarray}
\begin{eqnarray}
	V^{(n)} =
	\frac{1}{2 \sqrt{2}} 
	\int_0^\infty dr r^2 
	(A-Z) \rho^{(n)} ( g_e^- g_\mu^- + f_e^- f_\mu^- )
	\ ,
	\label{26}
\end{eqnarray}
where the functions $g_e^-$ etc.\ are defined in the $1s$ muon wave
function and $\kappa = \pm 1$ electron wave functions as follows:
\begin{eqnarray}
	\psi^{(\mu)}_{1s} (r,\theta,\phi) = \left(
	\begin{array}{c}
		g^-_{\mu} (r) \chi^{\pm 1/2}_{-1} (\theta, \phi)\\
		if^-_{\mu} (r) \chi^{\pm 1/2}_{1} (\theta, \phi)\\
	\end{array}
	\right),
	\label{9}
\end{eqnarray}
\begin{eqnarray}
	\psi^{\mu = \pm 1/2 (e)}_{\kappa=-1, W} (r,\theta,\phi) = 
	\left(
	\begin{array}{c}
		g^-_{e} (r) \chi^{\pm 1/2}_{-1} (\theta, \phi)\\
		if^-_{e} (r) \chi^{\pm 1/2}_{1} (\theta, \phi)\\
	\end{array}
	\right)\ ,
\end{eqnarray}
\begin{eqnarray}
	\psi^{\mu = \pm 1/2 (e)}_{\kappa=1, W} (r,\theta,\phi) = 
	\left(
	\begin{array}{c}
		g^+_{e} (r) \chi^{\pm 1/2}_{1} (\theta, \phi)\\
		if^+_{e} (r) \chi^{\pm 1/2}_{-1} (\theta, \phi)\\
	\end{array}
	\right)\ .
	\label{10}
\end{eqnarray}
In the above expression, we have neglected the electron mass, so that
$g_e^+$ and $f_e^+$ are related to $g_e^-$ and $f_e^-$ as $g_e^+ = i
f_e^-$ and $i f_e^+ = g_e^-$.
By integrating the Maxwell equations, the electric field $E(r)$ is
determined as
\begin{equation}
	E(r) = \frac{Ze}{r^2} \int_0^r r^{\prime 2} 
	\rho^{(p)}(r^\prime) dr^\prime\ .
	\label{eq:e-field}
\end{equation}

Once the proton and neutron densities are given, one can calculate the
electric field $E(r)$ by Eq.(\ref{eq:e-field}) and the electric
potential $V(r)$ by
\begin{equation}
	V(r) = -e \int_r^\infty E(r^\prime) dr^\prime\ .
	\label{eq:e-potential}
\end{equation}
The wave functions are then obtained by the Dirac equation
Eq.(\ref{5}), and the $\mu$-$e$ conversion rate is calculated by
Eq.(\ref{eq:omega-final}).

\section{Numerical results}
\label{section:results}
In order to evaluate Eq.(\ref{eq:omega-final}), we need proton and
neutron densities and the muon and electron wave functions.
We first discuss proton and neutron densities and show the wave
functions of muon and electron.
Then, we present numerical results of the overlap integrals
Eqs.(\ref{22}) -- (\ref{26}) and the conversion rates for various
nuclei.
%


\subsection{Distribution of protons and neutrons in the nucleus}
\label{subsection:nucleon-in-nucleus}

We have used proton densities determined from electron scattering
experiments.
In the past, the charge density distribution of a nucleus was analyzed
assuming some functional form of the proton distribution such as the
two-parameter Fermi model, the three-parameter Fermi model, and the
three-parameter Gaussian model.
More recently, with improvement of experimental data,
model-independent methods are used to extract detailed information on
the density distribution.  Examples are the Fourier-Bessel expansion
and the sum of Gaussian functions.
The proton density is very precisely determined when the
model-independent analysis is carried out.  We use the charge density
listed in Ref.~\cite{table}.  We adopt the results of the
model-independent analysis when the data are available.  More on
charge distribution is described in Appendix
\ref{appendix:nucleon-densities}.

The determination of the neutron distribution is not as easy as that
of the proton distribution \cite{Batty}.
There are several ways to determine the neutron density in the 
different region.
Pionic atoms provide a probe of the neutron density on the periphery
of the nucleus.  In a certain level of the pionic atom the pion is
absorbed by the nucleus.  Since the strong interaction between the
pion and the nucleus changes the energy and the width of this level,
we can obtain information on the neutron density in the nucleus from
the analysis of the atomic X-ray spectrum.
Scattering experiments on the nucleus by strong interacting particles
such as the proton, the alpha particle, and the charged pion are also
useful to determine the neutron density.
Recently, an experiment with antiprotonic atoms was carried out to
determine the proton and neutron density distribution in the periphery
region of various nuclei \cite{Trzcinska:sy}.

In this paper we use the following three methods to evaluate the
$\mu$-$e$ conversion ratio:
\begin{enumerate}
    \item
	First, we take the proton density from electron scattering
	experiments given in Appendix \ref{appendix:nucleon-densities}
	and assume the neutron density is the same as the proton
	density.  For light nuclei this is a good approximation
	because the number of neutrons and protons are approximately
	equal and the conversion rates do not depend on the details of
	the neutron distribution.  (method \ref{method:p=n})
    \label{method:p=n}
    \item
	Second, we employ the neutron distribution obtained from the
	pionic atom.  We use the results of the analysis of
	Ref.~\cite{Garcia-Recio:wk}, where the proton and neutron
	distributions are given in terms of the two-parameter Fermi
	function.  (method \ref{method:pion})
    \label{method:pion}
    \item
	Finally, we use the neutron distribution obtained from the
	polarized proton scattering experiment.  The analysis was carried
	out for carbon (C) \cite{Ray:1978gz}, titanium (Ti)
	\cite{Pauletta:1981pc}, nickel (Ni) \cite{Ray:1978gz}, zirconium
	(Zr) \cite{Ray:1978gz}, and lead (Pb) \cite{Ray:1978gz} where
	proton and neutron density are given in the literature.
	We also estimate the uncertainty due to the error of the neutron
	distribution from the scattering experiment based on
	Refs.\cite{Ray:1979qv,Hoffmann:1980kg,Starodubsky:xt}.  (method
	\ref{method:proton})
    \label{method:proton}
\end{enumerate}
The first method gives precise evaluation for the overlap integrals
$D$, $S^{(p)}$ and $V^{(p)}$.  On the other hand, the neutron density is
necessary in order to determine $S^{(n)}$ and $V^{(n)}$.
Since both methods \ref{method:pion} and \ref{method:proton} use strong
interacting particles as probes to the neutron density, the calculation
suffers from ambiguities associated with the pion/proton-nucleon
interaction.
The method \ref{method:pion} provides information on the size of the
nucleus in a wide range of atomic numbers.  On the other
hand, we can determine in the method \ref{method:proton} the profile of
the neutron distribution inside the nucleus for selected nuclei.
We calculate the conversion rate in two methods, and first discuss
ambiguities in each method and then compare the results.


\subsection{Numerical evaluation of the overlap integrals}
\label{subsec:results}

In this subsection, we first show an example of the muon and electron
wave functions obtained by solving the Dirac equation Eq.(\ref{2}),
and present the result of numerical calculation of overlap integrals
$D$, $S^{(p,n)}$ and $V^{(p,n)}$ defined by Eqs.(\ref{22}) --
(\ref{26}).


\subsubsection{Wave functions of the initial and final states}
\label{subsubsec:wavefunctions}
\begin{figure}[t]
    \begin{center}
	\includegraphics[width=15cm]{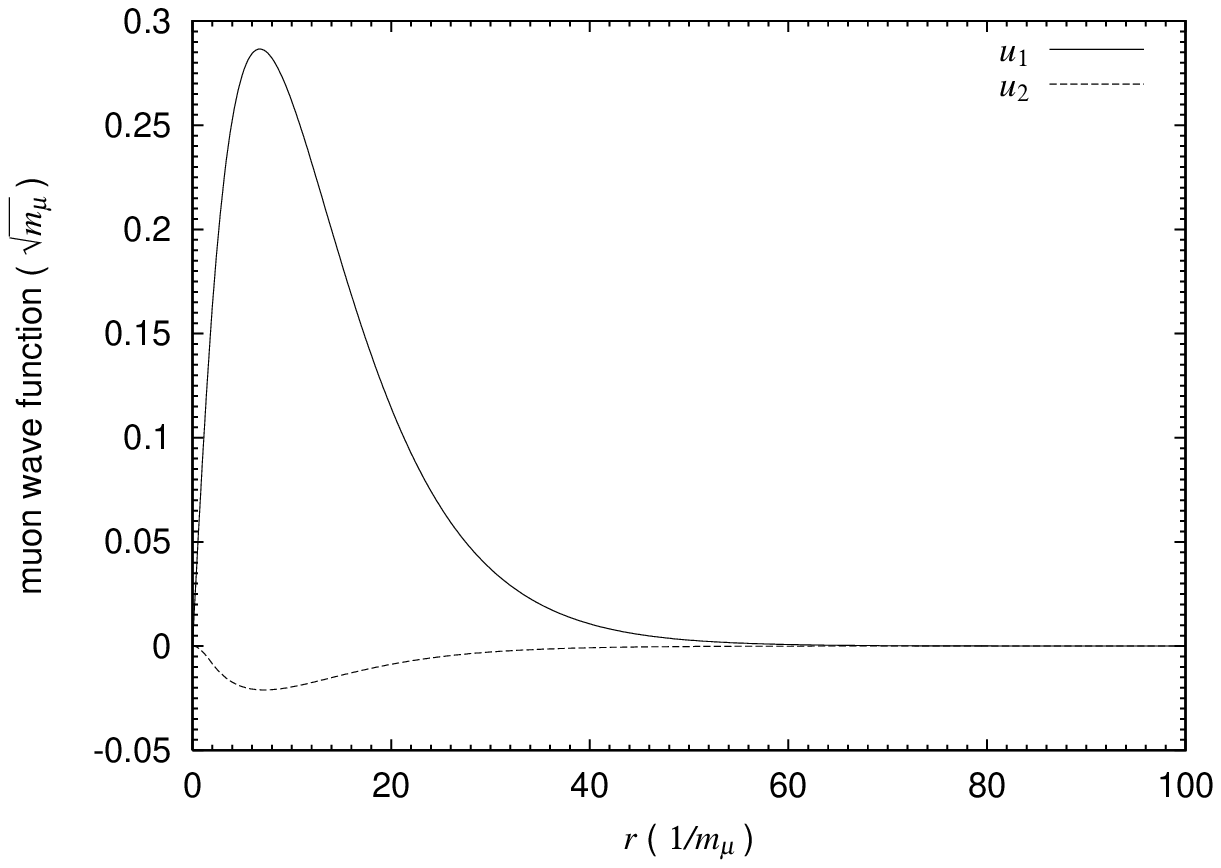} 
    \end{center}
    \caption{%
	The normalized wave function of a muon in the titanium (Ti)
	nucleus is plotted.  The solid line and dashed line represent
	$u_1 ( \equiv r g_\mu^- )$ and $u_2 (\equiv r f_\mu^-)$
	components, respectively.  The horizontal axis is the distance
	between the nucleus and the muon in the unit of $1/m_\mu$. 
	The unit for the wave function is
    taken to be $\sqrt{m_\mu}$.%
    }
    \label{fig1}
\end{figure}
\begin{figure}[t]
    \begin{center}
	\includegraphics[width=15cm]{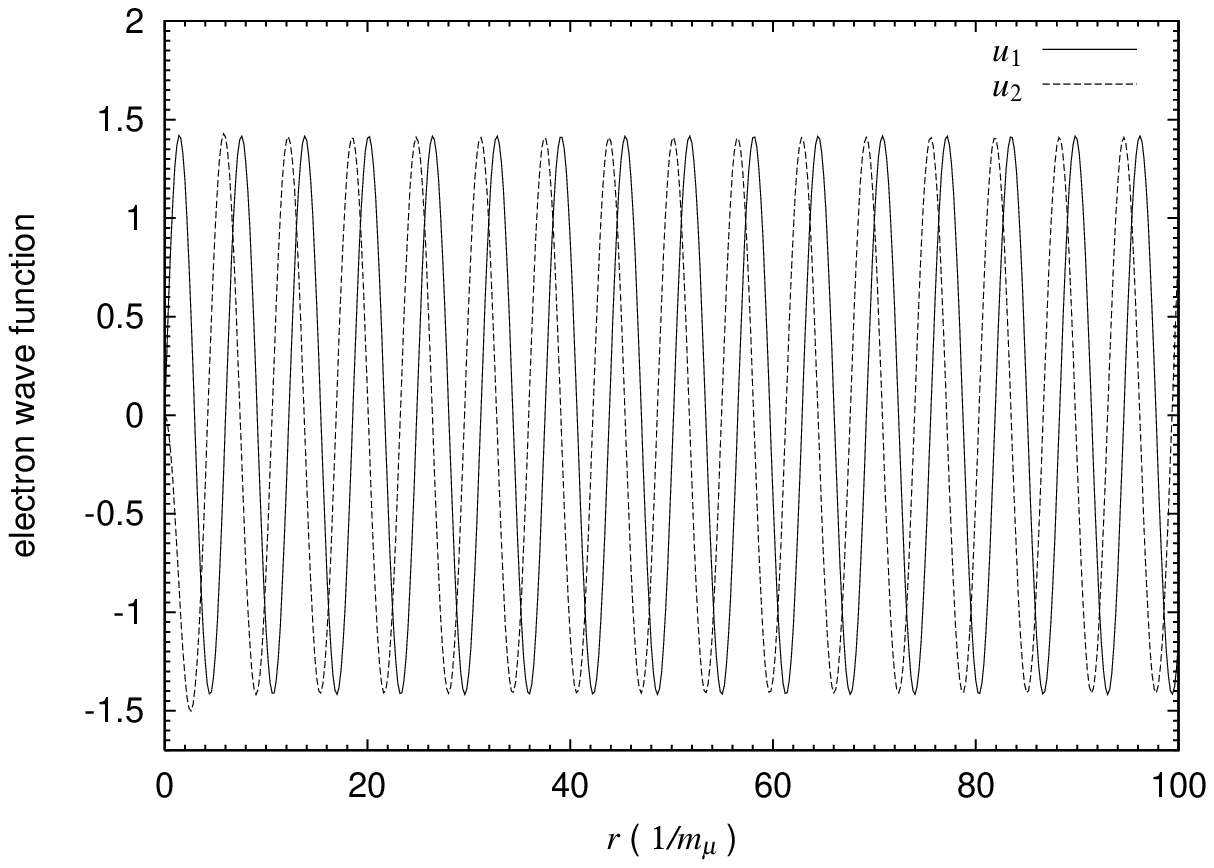} 
    \end{center}
    \caption{%
	The normalized wave function of an electron in titanium (Ti)
	nucleus is plotted.  The solid line and dashed line represent $u_1
	( \equiv r g_e^- )$ and $u_2 (\equiv r f_e^-)$ components,
	respectively.  The horizontal axis is the distance between the
	nucleus and the
    electron in the unit of $1/m_\mu$.%
    }
    \label{fig2}
\end{figure}

The muon and electron wave functions are evaluated by solving
the Dirac equation (\ref{2}) with the electric potential given by 
Eq.(\ref{eq:e-potential}).
Ignoring the recoil of the nucleus which is of the order of $m_\mu^2 /
M_N$, where $M_N$ is the nucleus mass, one finds that the energy of
the out-going electron in Eq.(\ref{2}) is equal to the muon mass minus
the binding energy.
As an example, we show the muon and electron wave functions in
titanium (Ti) nucleus in Figs.\ref{fig1} and \ref{fig2}.
We can see in Fig.\ref{fig1} that the lower component $u_2$ in the
muon wave function is much smaller than the upper component $u_1$.
However, as pointed out in Ref.~\cite{Shanker:1979ap},
its effect on the conversion rate is sizable for heavy nuclei.
The overlap integrals are evaluated using these wave functions.


\subsubsection{Method 1}
\label{subsubsec:method1}

\begin{table}[p]
    \begin{center}
	\begin{tabular}{|c||l|l|l|l|l|}
	    \hline
	    Nucleus & $D$ & $S^{(p)}$ & $V^{(p)}$ 
	    & $S^{(n)}$ & $V^{(n)}$ 
	    \\ 
	    \hline \hline
	    $^{4}_{2}$He & 0.000625 & 0.000262 & 0.000263 & 0.000262 & 0.000263 \\
$^{7}_{3}$Li & 0.00138 & 0.000581 & 0.000585 & 0.000775 & 0.000780 \\
$^{9}_{4}$Be & 0.00268 & 0.00113 & 0.00114 & 0.00141 & 0.00142 \\
$^{11}_{5}$B & 0.00472 & 0.00200 & 0.00202 & 0.00240 & 0.00242 \\
$^{12}_{6}$C & 0.00724 & 0.00308 & 0.00312 & 0.00308 & 0.00312 \\
$^{14}_{7}$N & 0.0103 & 0.0044 & 0.0044 & 0.0044 & 0.0044 \\
$^{16}_{8}$O & 0.0133 & 0.0057 & 0.0058 & 0.0057 & 0.0058 \\
$^{19}_{9}$F & 0.0166 & 0.0071 & 0.0072 & 0.0079 & 0.0081 \\
$^{20}_{10}$Ne & 0.0205 & 0.0088 & 0.0090 & 0.0088 & 0.0090 \\
$^{24}_{12}$Mg & 0.0312 & 0.0133 & 0.0138 & 0.0133 & 0.0138 \\
$^{27}_{13}$Al & 0.0362 & 0.0155 & 0.0161 & 0.0167 & 0.0173 \\
$^{28}_{14}$Si & 0.0419 & 0.0179 & 0.0187 & 0.0179 & 0.0187 \\
$^{31}_{15}$P & 0.0468 & 0.0201 & 0.0210 & 0.0214 & 0.0224 \\
$^{32}_{16}$S & 0.0524 & 0.0225 & 0.0236 & 0.0225 & 0.0236 \\
$^{35}_{17}$Cl & 0.0565 & 0.0241 & 0.0254 & 0.0256 & 0.0269 \\
$^{40}_{18}$Ar & 0.0621 & 0.0265 & 0.0281 & 0.0324 & 0.0343 \\
$^{39}_{19}$K & 0.0699 & 0.0299 & 0.0317 & 0.0314 & 0.0334 \\
$^{40}_{20}$Ca & 0.0761 & 0.0325 & 0.0347 & 0.0325 & 0.0347 \\
$^{48}_{22}$Ti & 0.0864 & 0.0368 & 0.0396 & 0.0435 & 0.0468 \\
$^{51}_{23}$V & 0.0931 & 0.0396 & 0.0428 & 0.0482 & 0.0521 \\
$^{52}_{24}$Cr & 0.100 & 0.0425 & 0.0461 & 0.0496 & 0.0538 \\
$^{55}_{25}$Mn & 0.107 & 0.0456 & 0.0496 & 0.0547 & 0.0596 \\
$^{56}_{26}$Fe & 0.110 & 0.0467 & 0.0512 & 0.0539 & 0.0591 \\
$^{59}_{27}$Co & 0.112 & 0.0471 & 0.0519 & 0.0558 & 0.0615 \\
$^{58}_{28}$Ni & 0.125 & 0.0527 & 0.0583 & 0.0565 & 0.0625 \\
$^{63}_{29}$Cu & 0.122 & 0.0514 & 0.0572 & 0.0603 & 0.0671 \\
$^{64}_{30}$Zn & 0.134 & 0.0561 & 0.0627 & 0.0636 & 0.0710 \\
$^{74}_{32}$Ge & 0.133 & 0.0554 & 0.0628 & 0.0727 & 0.0824 \\
$^{80}_{34}$Se & 0.146 & 0.0602 & 0.0690 & 0.0815 & 0.0933 \\
$^{88}_{38}$Sr & 0.163 & 0.0665 & 0.0778 & 0.0875 & 0.102 \\
$^{89}_{39}$Y & 0.164 & 0.0670 & 0.0788 & 0.0859 & 0.101 \\

	    \hline 
	\end{tabular}
    \end{center}
    \caption{ The overlap integrals in the unit of $m_\mu^{5/2}$ are
    listed.
    The proton distribution in the nuclei are taken from
    Ref.~\cite{table} (see also Appendix
    \ref{appendix:nucleon-densities}), and neutron distribution are
    assumed to be same as that of protons (method \ref{method:p=n} in
    subsection \ref{subsection:nucleon-in-nucleus}).  } \label{tab1}
\end{table}
\addtocounter{table}{-1}
\begin{table}[htp]
    \begin{center}
	\begin{tabular}{|c||l|l|l|l|l|}
	    \hline
	    Nucleus &
	    $D$ & $S^{(p)}$ & $V^{(p)}$ 
	    & $S^{(n)}$ & $V^{(n)}$ 
	    \\ 
	    \hline \hline
	    $^{90}_{40}$Zr & 0.171 & 0.0697 & 0.0823 & 0.0872 & 0.103 \\
$^{93}_{41}$Nb & 0.171 & 0.0692 & 0.0823 & 0.0878 & 0.104 \\
$^{98}_{42}$Mo & 0.170 & 0.0683 & 0.0818 & 0.0911 & 0.109 \\
$^{110}_{46}$Pd & 0.176 & 0.0695 & 0.0855 & 0.0967 & 0.119 \\
$^{114}_{48}$Cd & 0.178 & 0.0696 & 0.0867 & 0.0958 & 0.119 \\
$^{115}_{49}$In & 0.181 & 0.0704 & 0.0882 & 0.0948 & 0.119 \\
$^{120}_{50}$Sn & 0.183 & 0.0707 & 0.0894 & 0.0990 & 0.125 \\
$^{121}_{51}$Sb & 0.195 & 0.0760 & 0.0957 & 0.104 & 0.131 \\
$^{138}_{56}$Ba & 0.184 & 0.0688 & 0.0911 & 0.101 & 0.133 \\
$^{139}_{57}$La & 0.189 & 0.0707 & 0.0937 & 0.102 & 0.135 \\
$^{142}_{60}$Nd & 0.183 & 0.0669 & 0.0909 & 0.0914 & 0.124 \\
$^{152}_{62}$Sm & 0.175 & 0.0631 & 0.0875 & 0.0915 & 0.127 \\
$^{158}_{64}$Gd & 0.173 & 0.0613 & 0.0865 & 0.0901 & 0.127 \\
$^{165}_{67}$Ho & 0.177 & 0.0617 & 0.0892 & 0.0902 & 0.131 \\
$^{166}_{68}$Er & 0.200 & 0.0693 & 0.101 & 0.0999 & 0.146 \\
$^{181}_{73}$Ta & 0.156 & 0.0513 & 0.0792 & 0.0759 & 0.117 \\
$^{184}_{74}$W & 0.156 & 0.0499 & 0.0794 & 0.0741 & 0.118 \\
$^{197}_{79}$Au & 0.189 & 0.0614 & 0.0974 & 0.0918 & 0.146 \\
$^{204}_{80}$Hg & 0.158 & 0.0482 & 0.0818 & 0.0746 & 0.127 \\
$^{205}_{81}$Tl & 0.161 & 0.0491 & 0.0834 & 0.0752 & 0.128 \\
$^{208}_{82}$Pb & 0.161 & 0.0488 & 0.0834 & 0.0749 & 0.128 \\
$^{209}_{83}$Bi & 0.159 & 0.0475 & 0.0826 & 0.0722 & 0.125 \\
$^{232}_{90}$Th & 0.154 & 0.0429 & 0.0809 & 0.0677 & 0.128 \\
$^{238}_{92}$U & 0.151 & 0.0417 & 0.0798 & 0.0662 & 0.127 \\

	    \hline
	\end{tabular}
    \end{center}
    \caption{
    (Continued).
    }
\end{table}

\begin{figure}[t]
    \begin{center}
	\includegraphics[width=15cm]{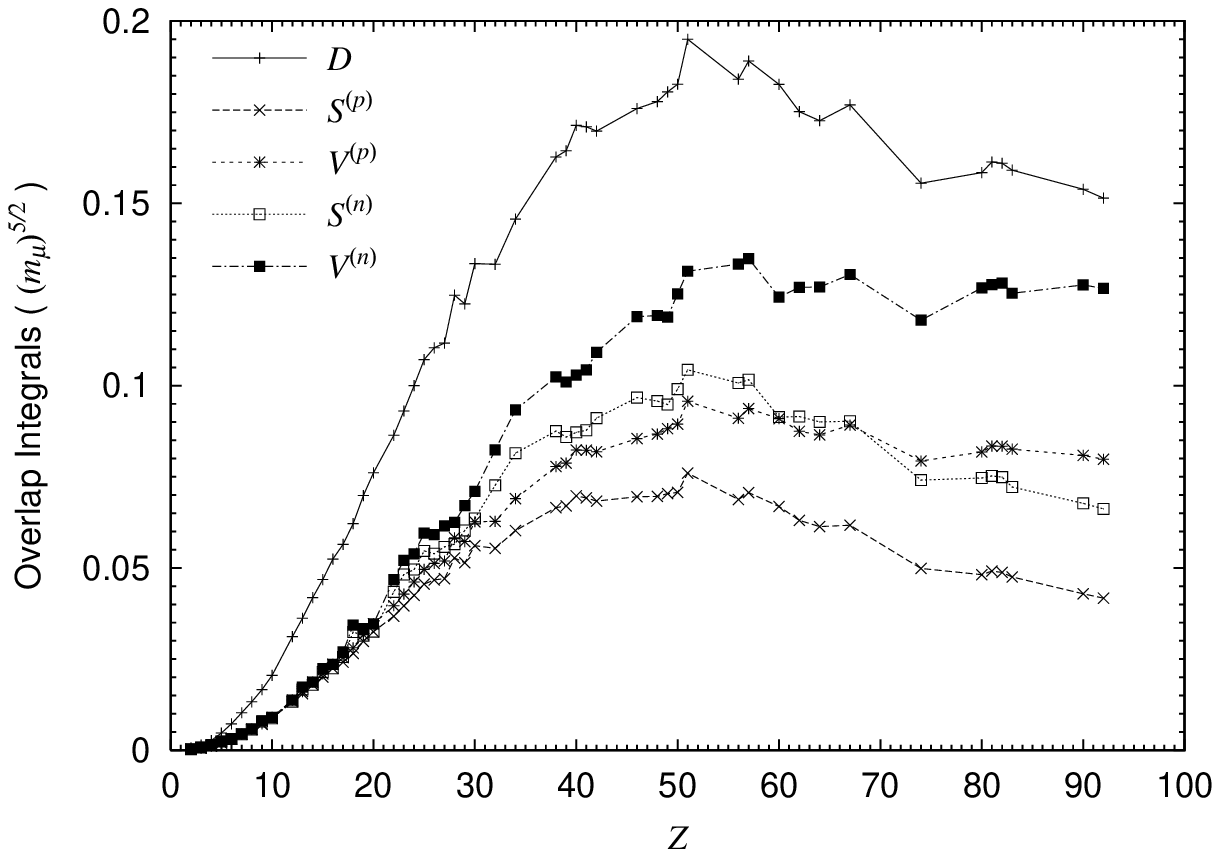} 
    \end{center}
\caption{ The overlap integrals are plotted as functions of the atomic
number $Z$. Neutron density distributions are assumed to be equal to
charge density distributions (method \ref{method:p=n} in subsection
\ref{subsection:nucleon-in-nucleus}). } \label{fig:Overlap-Zdep}
\end{figure}

We present in Table \ref{tab1} the results under the assumption
$\rho_{n} = \rho_{p}$, where $\rho_{p}$ is taken from Ref.~\cite{table}.
We show in Fig.\ref{fig:Overlap-Zdep} the $Z$ dependence of the
integrals.  We omitted the points for $^{166}_{68}$Er, $^{181}_{73}$Ta,
and $^{197}_{79}$Au from this figure since these data are either
obtained from quite old experiments or not well-established
\cite{table}.
We see that the overlap integrals increase as functions of $Z$
for light nuclei up to $Z \sim 30$, and saturate or decrease for heavy
nuclei.

Although the tendency is the same, each overlap integral has different
$Z$ dependence, especially for heavy nuclei.
For example, the scalar ($S^{(p,n)}$) and the vector ($V^{(p,n)}$)
type integrals are almost equal in light nuclei $( Z \lesssim 30 )$,
whereas the vector like integral is larger by a factor of 1.5 -- 2
than the scalar one for heavy nuclei.
This difference is due to the the relativistic effects of the muon
wave functions which are significant in heavy nuclei.
In fact, the scalar and vector overlap integrals in Eqs.(\ref{23}) and
(\ref{25}) [Eqs.(\ref{24}) and (\ref{26})] are exactly the same, if we
ignore the small component of the wave function $f_\mu^-$.
For $D$ we can show that $D / (8e) \simeq S^{(p)} \simeq V^{(p)}$ is
satisfied for light nuclei.

\begin{figure}[t]
    \begin{center}
	\includegraphics[width=15cm]{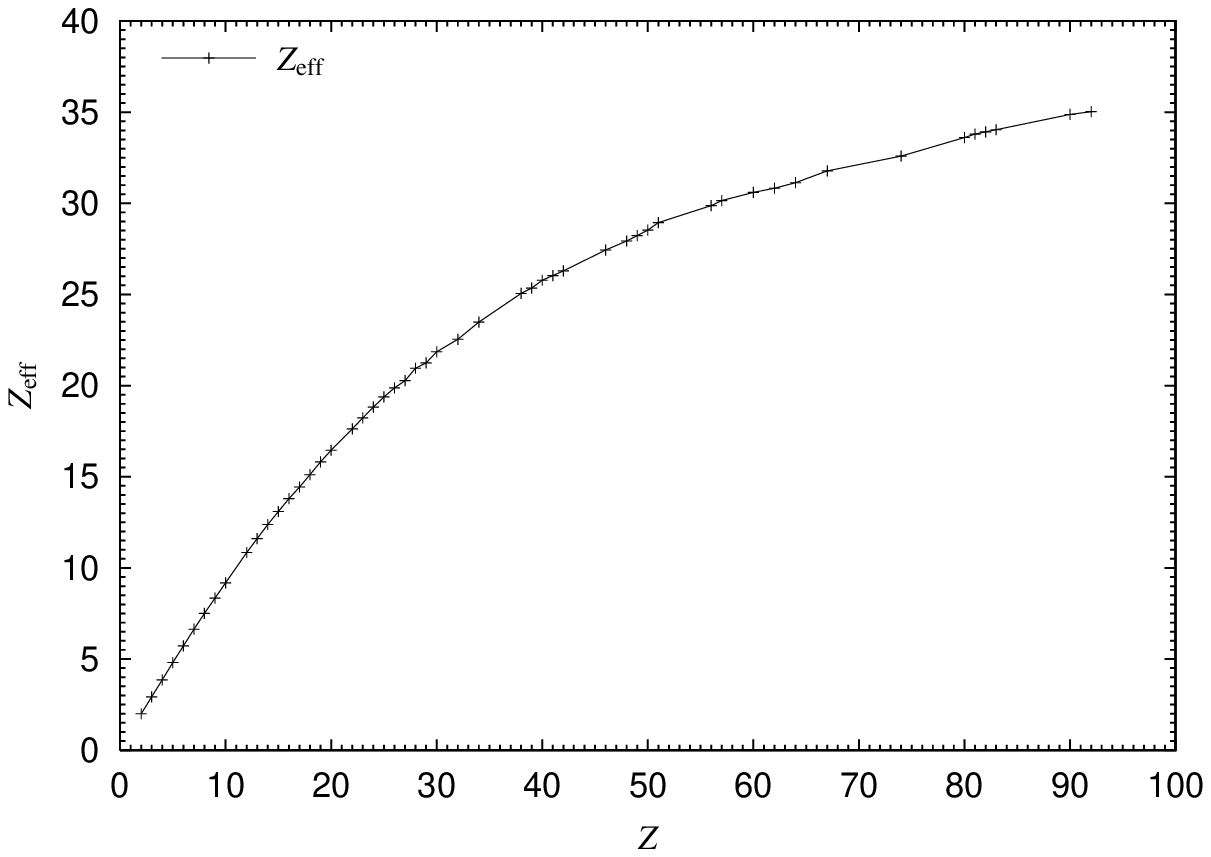} 
    \end{center}
    \caption{%
    The effective charge for the muon in the $1s$ state is plotted as
    a function of the atomic number $Z$.%
    }
    \label{fig4}
\end{figure}

\begin{figure}[t]
    \begin{center}
	\includegraphics[width=15cm]{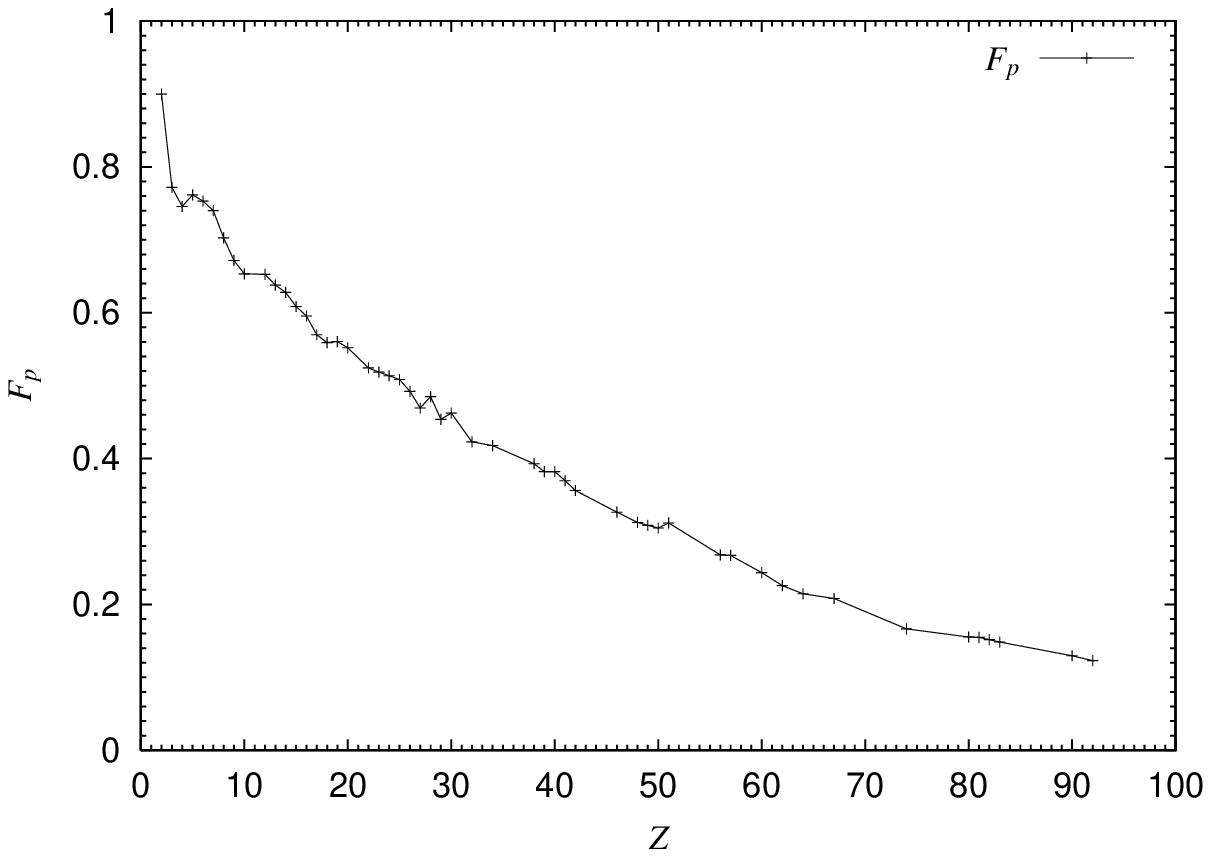} 
    \end{center}
    \caption{%
    The form factor $F_{p}$ is plotted as a function of the
    atomic number $Z$.%
    } \label{fig5}
\end{figure}

The qualitative feature of the $Z$ dependence of the overlap integrals
can be understood from the following consideration.
When we adopt an approximation by Weinberg and Feinberg \cite{WF}, where
the muon wave function is replaced by the average value and the electron
wave function is treated as a plane wave, the formula for the overlap
integrals in Eqs.(\ref{23}) and (\ref{25}) are given by
\begin{equation}
    V^{(p)} \sim
	S^{(p)} \sim
    \frac{1}{8\pi} \langle \phi_\mu \rangle Z F_p\ .
\end{equation}
Here $F_{p}$ is the form factor defined by
\begin{equation}
    F_{p} =
	\int_0^\infty
	dr 4 \pi r^2 \rho^{(p)}\ 
    \frac{\sin m_\mu r}{m_\mu r}\ ,
\end{equation}
and
$\langle \phi_\mu \rangle$ is the average value
of the muon wave function in the nucleus
given by
\begin{eqnarray}
    \langle \phi_\mu \rangle^2 =
    \int_0^\infty dr 4 \pi r^2 \ (g_\mu^2 + f_\mu^2) \ \rho^{(p)}
    = \frac{4 m_\mu^3 \alpha^3 Z_{\rm eff}^4}{ Z }\ .
\end{eqnarray}
In the last expression, we have introduced $Z_{\rm eff}$ which is the
effective charge for the muon in the $1s$ state.
We show $Z_{\rm eff}$ in Fig.\ref{fig4}.
Since the muon wave function enter into
the inside of the nucleus,
$Z_{\rm eff}$ thus does not increase
linearly with respect to $Z$.
In Fig.\ref{fig5}, we show the form factor $F_p$ calculated based on
method \ref{method:p=n}.  There we see that $F_p$ is a decreasing
function and suppressed for heavy nuclei.  These two properties
explain the $Z$ dependence of the overlap integrals.

%


\subsubsection{Method 2}
\label{subsubsec:method2}
\begin{figure}[t]
    \begin{center}
	\includegraphics[width=15cm]{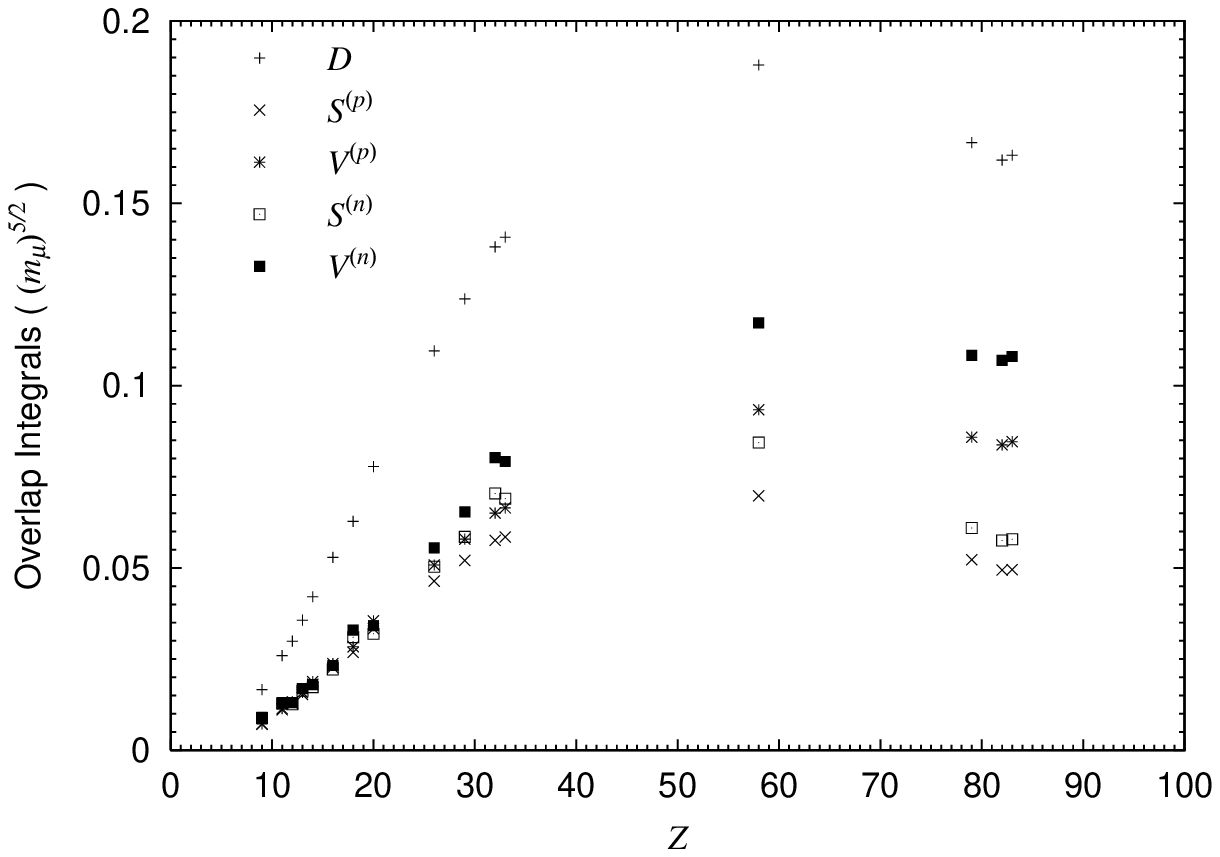} 
    \end{center}
    \caption{%
	Same as Fig.\ref{fig:Overlap-Zdep}, but here the results of
	the analysis of the pionic atom experiment are used for the
	distribution of the neutrons \cite{Garcia-Recio:wk} (method
	\ref{method:pion} in subsection
	\ref{subsection:nucleon-in-nucleus}).%
    }
    \label{fig:Overlap-pion}
\end{figure}

We present in Table \ref{table:Overlap-pion} and
Fig.\ref{fig:Overlap-pion} the values of the overlap integrals in the
method \ref{method:pion}, in which the input nucleon distributions are
obtained from the analysis of the pionic atom experiments
\cite{Garcia-Recio:wk}.
In this method, the proton and neutron density are assumed to be a
two-parameter Fermi model defined by Eq.(\ref{eq:2pF-def}) in Appendix
\ref{appendix:nucleon-densities}.
Using the energy shift and the decay width of the pionic atom, the
neutron density is determined together with parameters in the
pion-nucleon optical potential.  In Ref.\cite{Garcia-Recio:wk}, it is
assumed that the diffuseness of the neutron density is the same as
that of the proton density, $z_{n} = z_{p}$, so that the output
parameter is the radius of the neutron density.
For our calculation, we use the neutron matter parameter $R_{n}$[mean]
in Table 4 of Ref.~\cite{Garcia-Recio:wk}.
The $Z$ dependence shown in Fig.\ref{fig:Overlap-pion} is similar to
that seen in Fig.\ref{fig:Overlap-Zdep}.

\begin{table}[pth]
    \begin{center}
	\begin{tabular}{|c||l|l|l|l|l|}
	    \hline
	    Nucleus &
	    $D$ & $S^{(p)}$ & $V^{(p)}$ 
	    & $S^{(n)}$ & $V^{(n)}$ 
	    \\ 
	    \hline \hline
	    $^{19}_{9}$F & 0.0166 & 0.0071 & 0.0072 & 0.0089 & 0.0090 \\
$^{23}_{11}$Na & 0.0260 & 0.0111 & 0.0114 & 0.0128 & 0.0131 \\
$^{24}_{12}$Mg & 0.0299 & 0.0128 & 0.0132 & 0.0126 & 0.0131 \\
$^{27}_{13}$Al & 0.0357 & 0.0153 & 0.0159 & 0.0163 & 0.0169 \\
$^{28}_{14}$Si & 0.0421 & 0.0181 & 0.0188 & 0.0173 & 0.0180 \\
$^{32}_{16}$S & 0.0529 & 0.0227 & 0.0238 & 0.0221 & 0.0233 \\
$^{40}_{18}$Ar & 0.0628 & 0.0268 & 0.0284 & 0.0310 & 0.0330 \\
$^{40}_{20}$Ca & 0.0778 & 0.0333 & 0.0355 & 0.0319 & 0.0341 \\
$^{56}_{26}$Fe & 0.110 & 0.0464 & 0.0508 & 0.0503 & 0.0555 \\
$^{63}_{29}$Cu & 0.124 & 0.0521 & 0.0579 & 0.0585 & 0.0654 \\
$^{74}_{32}$Ge & 0.138 & 0.0576 & 0.0651 & 0.0704 & 0.0802 \\
$^{75}_{33}$As & 0.141 & 0.0585 & 0.0665 & 0.0690 & 0.0792 \\
$^{142}_{58}$Ce & 0.188 & 0.0698 & 0.0934 & 0.0844 & 0.117 \\
$^{197}_{79}$Au & 0.167 & 0.0523 & 0.0859 & 0.0610 & 0.108 \\
$^{208}_{82}$Pb & 0.162 & 0.0495 & 0.0838 & 0.0575 & 0.107 \\
$^{209}_{83}$Bi & 0.163 & 0.0495 & 0.0846 & 0.0579 & 0.108 \\

	    \hline 
	\end{tabular}
    \end{center}
    \caption{ Same as Table \ref{tab1}, but here the results of the
    analysis of the pionic atom experiment are used for the distribution
    of the neutrons \cite{Garcia-Recio:wk} (method \ref{method:pion} in
    subsection \ref{subsection:nucleon-in-nucleus}). } 
    \label{table:Overlap-pion}
\end{table}

In order to estimate the ambiguity within this method, we have
calculated $S^{(n)}$ and $V^{(n)}$ by changing the radius of the
neutron density $c_{n}$ within the error given in the
Ref.\cite{Garcia-Recio:wk}.
We show the results for $^{27}_{13}$Al, $^{142}_{58}$Ce, and
$^{208}_{82}$Pb corresponding to light, intermediate, and heavy nuclei
in Table \ref{table:Pb-ambiguity}.
The error of $c_n$ includes the statistical error as well as the
systematic error estimated from the outputs in two different
parameterizations of the optical potential \cite{Garcia-Recio:wk}.
We see that $S^{(n)}$ and $V^{(n)}$ change $\pm 2\%$ for $^{27}_{13}$Al,
$\pm (6-8)\%$ for $^{142}_{58}$Ce, and $\pm (7-11)\%$ for
$^{208}_{82}$Pb.

\begin{table}[t]
    \begin{center}
	\begin{tabular}{|c||l|l|l|}
	    \hline
	    Nucleus & $c_n$ [fm] &  $S^{(n)}$ & $V^{(n)}$ \\
	    \hline \hline
$^{27}_{13}$Al & 3.09 $\pm$ 0.08  & 0.0163 $\mp$ 0.003 & 0.0169 $\mp$ 0.003 \\
$^{142}_{58}$Ce & 6.00 $\pm$ 0.10  & 0.0844 $\mp$ 0.0067 & 0.117 $\mp$ 0.008 \\
$^{208}_{82}$Pb & 6.86 $\pm$ 0.09  & 0.0575 $\mp$ 0.0066 & 0.107 $\mp$ 0.008 \\
	    \hline
	\end{tabular}
    \end{center}
    \caption{%
    Error of the overlap integrals associated with the input
    value of the neutron radii determined by the pionic atom experiment.
    The larger (smaller) values of $S^{(n)}$ and $V^{(n)}$ correspond to
    smaller (larger) value of $c_n$.%
    }
    \label{table:Pb-ambiguity}
\end{table}


\subsubsection{Method 3}
\label{subsubsec:method3}
The neutron densities for selected nuclei were determined from the 800
MeV polarized proton elastic scattering experiments performed at LAMPF
in the late 1970's.  In Refs.\cite{Ray:1978gz,Pauletta:1981pc}, proton
and neutron densities are given assuming the three-parameter Fermi or
Gaussian model.
Errors of the neutron distributions are estimated in a model
independent fashion in Refs.\cite{Ray:1979qv,Hoffmann:1980kg}.
More recently, the determination of the neutron density have been
improved for lead (Pb) based on a new experiment and a new analysis
\cite{Starodubsky:xt,Mack}.
\begin{figure}[t]
    \begin{center}
	\includegraphics[width=15cm]{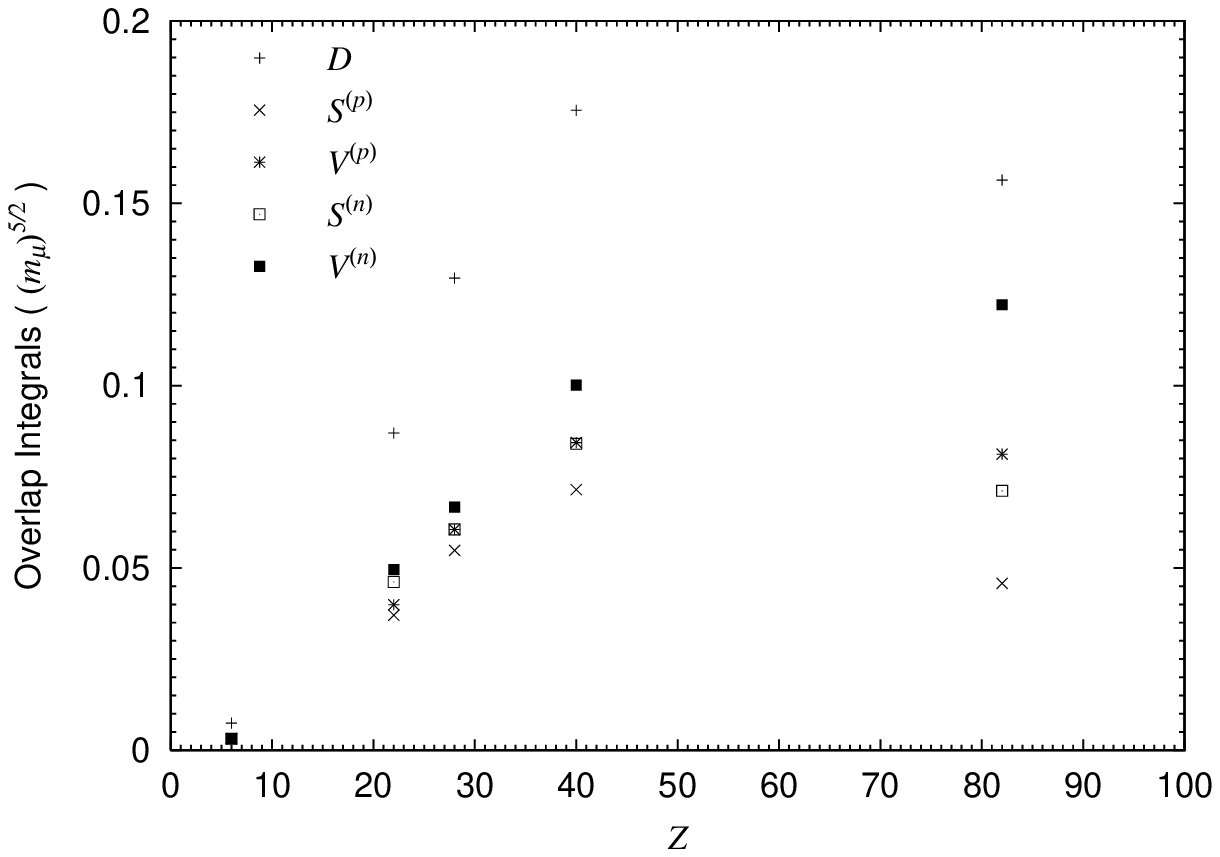} 
    \end{center}
    \caption{%
	Same as Fig.\ref{fig:Overlap-Zdep}, but here the results of the
	analysis of the proton scattering experiments are used for the
	neutron density distribution \cite{Ray:1978gz,Pauletta:1981pc}
	(method \ref{method:proton} in subsection
    \ref{subsection:nucleon-in-nucleus}).%
    }
    \label{fig:Overlap-proton}
\end{figure}

\begin{figure}[t]
    \begin{center}
	\includegraphics[width=15cm]{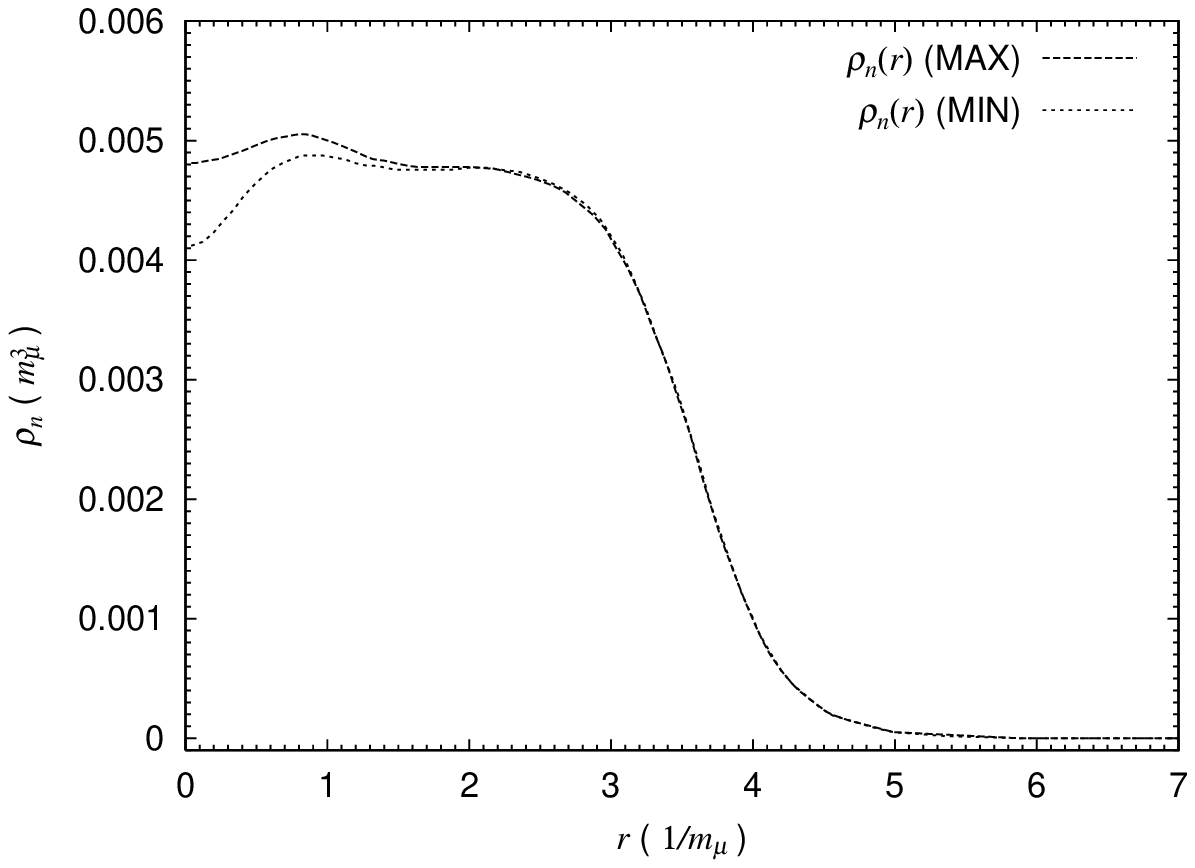} 
    \end{center}
    \caption{%
    The error envelope of the neutron density distribution for 
    $^{208}_{82}$Pb nucleus \cite{Starodubsky:xt}.%
    }
    \label{fig:Pb-density}
\end{figure}

We calculate the overlap integrals using neutron distribution given in
the above references.  Table \ref{table:Overlap-proton} and
Fig.\ref{fig:Overlap-proton} show the results for the five nuclei from
the Refs.\cite{Ray:1978gz,Pauletta:1981pc}.
In order to evaluate uncertainty from the neutron distribution
determined by the proton experiments, we take several examples where
the error of the neutron distribution is explicitly given in the
literature.
We calculated $S^{(n)}$ and $V^{(n)}$ within the uncertainty of the
neutron distribution.  For example, we present the error band of the
neutron distribution in the nucleus $^{208}_{82}$Pb in
Fig.\ref{fig:Pb-density} \cite{Starodubsky:xt}.
Similar error bands based on older experiments are given in
Ref.~\cite{Ray:1979qv} for $^{40}_{20}$Ca, $^{58}_{28}$Ni, and
$^{116}_{50}$Sn, and in Ref.~\cite{Hoffmann:1980kg} for
$^{208}_{82}$Pb.
The overlap integrals evaluated using the minimum and maximum values of
the envelope are shown in Tables \ref{table:Overlap-envelope} and
\ref{table:Overlap-envelope2}.
Since the proton distribution is not explicitly given except the case
of $^{208}_{82}$Pb in Ref.\cite{Starodubsky:xt}, we use the proton
distribution listed in Table \ref{tab_nucl} for other cases.
The errors of $S^{(n)}$ and $V^{(n)}$ amount to a few percent for the
light nuclei such as $^{40}_{20}$Ca and $^{58}_{28}$Ni.
For $^{208}_{82}$Pb, we can see drastic improvements in the
determination of $S^{(n)}$ and $V^{(n)}$ based on the new measurement. 
According to Ref.\cite{Starodubsky:xt}, the reduction of the errors in
the neutron distribution is mainly due to higher statistical accuracy
in the new experiment.
This implies that improvement in the proton scattering is important 
to determine neutron densities more precisely and reduce the ambiguity 
in the calculation of the $\mu$-$e$ conversion rate for heavy nuclei.

\begin{table}[ht]
    \begin{center}
	\begin{tabular}{|c||l|l|l|l|l|}
	    \hline
	    Nucleus &
	    $D$ & $S^{(p)}$ & $V^{(p)}$ 
	    & $S^{(n)}$ & $V^{(n)}$ 
	    \\ 
	    \hline \hline
	    $^{12}_{6}$C & 0.0074 & 0.0032 & 0.0032 & 0.0031 & 0.0032 \\
$^{48}_{22}$Ti & 0.0870 & 0.0371 & 0.0399 & 0.0462 & 0.0495 \\
$^{58}_{28}$Ni & 0.130 & 0.0548 & 0.0605 & 0.0606 & 0.0667 \\
$^{90}_{40}$Zr & 0.176 & 0.0715 & 0.0844 & 0.0841 & 0.100 \\
$^{208}_{82}$Pb & 0.156 & 0.0457 & 0.0812 & 0.0712 & 0.122 \\

	    \hline 
	\end{tabular}
    \end{center}
    \caption{ Same as Table \ref{tab1}, but here the results of the
    analysis of the proton scattering experiments are used for the
    neutron density distribution \cite{Ray:1978gz,Pauletta:1981pc}
    (method \ref{method:proton} in subsection
    \ref{subsection:nucleon-in-nucleus}). }
    \label{table:Overlap-proton}
\end{table}

\begin{table}[ht]
    \begin{center}
	\begin{tabular}{|cc||l|l|c|}
	    \hline
	    & & $S^{(n)}$ & $V^{(n)}$ & Ref. 
	    \\ 
	    \hline \hline
	    $^{40}_{20}$Ca
	    & Minimum & 0.0331  & 0.0352  & \cite{Ray:1979qv}\\
	    & Maximum & 0.0338  & 0.0359  & \\
	    \hline
	    $^{58}_{28}$Ni
	    & Minimum & 0.0584  & 0.0644  & \cite{Ray:1979qv} \\
	    & Maximum & 0.0592 & 0.0651 & \\
	    \hline
	    $^{116}_{50}$Sn
	    & Minimum & 0.0958 & 0.120   & \cite{Ray:1979qv} \\
	    & Maximum & 0.104  & 0.128  & \\
	    \hline
	    $^{208}_{82}$Pb
	    & Minimum & 0.0679 & 0.120  & \cite{Hoffmann:1980kg} \\
	    & Maximum & 0.0789 & 0.131  & \\	    %
	    \hline
	\end{tabular}
    \end{center}
    \caption{
	Maximum and minimum values of the overlap integrals for the
	neutron density distribution, which is changed within the error
	envelope. 
	}
    \label{table:Overlap-envelope}
\end{table}

\begin{table}[ht]
\begin{center}
 \begin{tabular}{|cc||l|l|l|l|l|c|}
\hline
  & & $D$ & $S^{(p)}$ & $V^{(p)}$ & $S^{(n)}$ & $V^{(n)}$ & Ref. \\
\hline \hline
  $^{208}_{82}$Pb & Minimum & 0.163 & 0.0493 & 0.0845 & 0.0675 & 0.0119
  & \cite{Starodubsky:xt} \\ 
  & Maximum & 0.163 & 0.0493 & 0.0845 & 0.0697 & 0.0121 & \\ \hline
 \end{tabular}
\end{center}
    \caption{ Maximum and minimum values of the overlap integrals for
    the neutron density distribution for $^{208}_{82}$Pb based on the
    analysis in Ref.~\cite{Starodubsky:xt}.  } 
    \label{table:Overlap-envelope2}
\end{table}


\subsubsection{Comparison of the results in three methods}

Comparing Tables \ref{tab1}, \ref{table:Overlap-pion}, and
\ref{table:Overlap-proton}, one finds that the overlap integrals of the
light nuclei agree with one another within a few percent.
For heavy nuclei, the method \ref{method:p=n} is not necessarily a good
approximation for $S^{(n)}$ and $V^{(n)}$.  However, we see that the
results are consistent with the values in the method \ref{method:proton}
within 10\%.
Comparing the method \ref{method:pion} and \ref{method:proton}, the
pionic atom method gives generally smaller values by $10 - 20 \%$ than
the analysis based on the proton scattering for the intermediate and
heavy nuclei.

We should note that the present analysis with pionic atoms assumes the
two-parameter Fermi model with $z_p = z_n$ for neutron distributions. 
In Ref.~\cite{Trzcinska:sy}, the neutron distribution is analyzed from
an antiprotonic atom experiment, namely the nuclear spectroscopy
analysis of the antiproton annihilation residues and the measurements
of strong-interaction effects on antiprotonic X-rays.
The authors compared two neutron distributions of neutron skin type
($z_n = z_p$ and $c_n > c_p$) and halo type ($z_n > z_p$ and $c_n =
c_p$), and concluded that the halo type distribution was favored.

In order to illustrate how the results depend on the assumption of the
neutron distribution, we calculate the $S^{(n)}$ and $V^{(n)}$ in the
halo type distribution with fixed values of the mean square radius.
For lead (Pb), $S^{(n)}$ $( V^{(n)} )$ increases by 22\% (12\%)
compared to the skin type distribution, so that the results are very
close to the values in the method \ref{method:proton}.
This indicates that the overlap integrals are sensitive to the neutron
distribution inside the nucleus, not only global quantities such as
the mean square radius.
In the method \ref{method:pion}, the pionic atom data provide us with
information on the neutron distribution mostly in the peripheral
region, and therefore the interior neutron distribution is given by an
extrapolation based on the assumed functional form.
Since the skin type distribution is assumed in
Ref.\cite{Garcia-Recio:wk}, there may be errors associated with this
assumption.

We have shown that in the method \ref{method:proton} the ambiguity of
$S^{(n)}$ and $V^{(n)}$ is reduced to a few percent even for heavy
nuclei such as lead (Pb) if we use the results of the new experiment.
The neutron distribution derived by experiments depend on how we treat
the scattering of a proton and a nucleus, so that there might be
uncertainties in the evaluation of $S^{(n)}$ and $V^{(n)}$ associated 
with validity of the scattering theory.
The improvement of the scattering theory would be necessary to give
more realistic error estimation \cite{Karataglidis:2001yn}.


\subsection{Numerical evaluation of conversion rate}

\begin{figure}[t]
    \begin{center}
	\includegraphics[width=15cm]{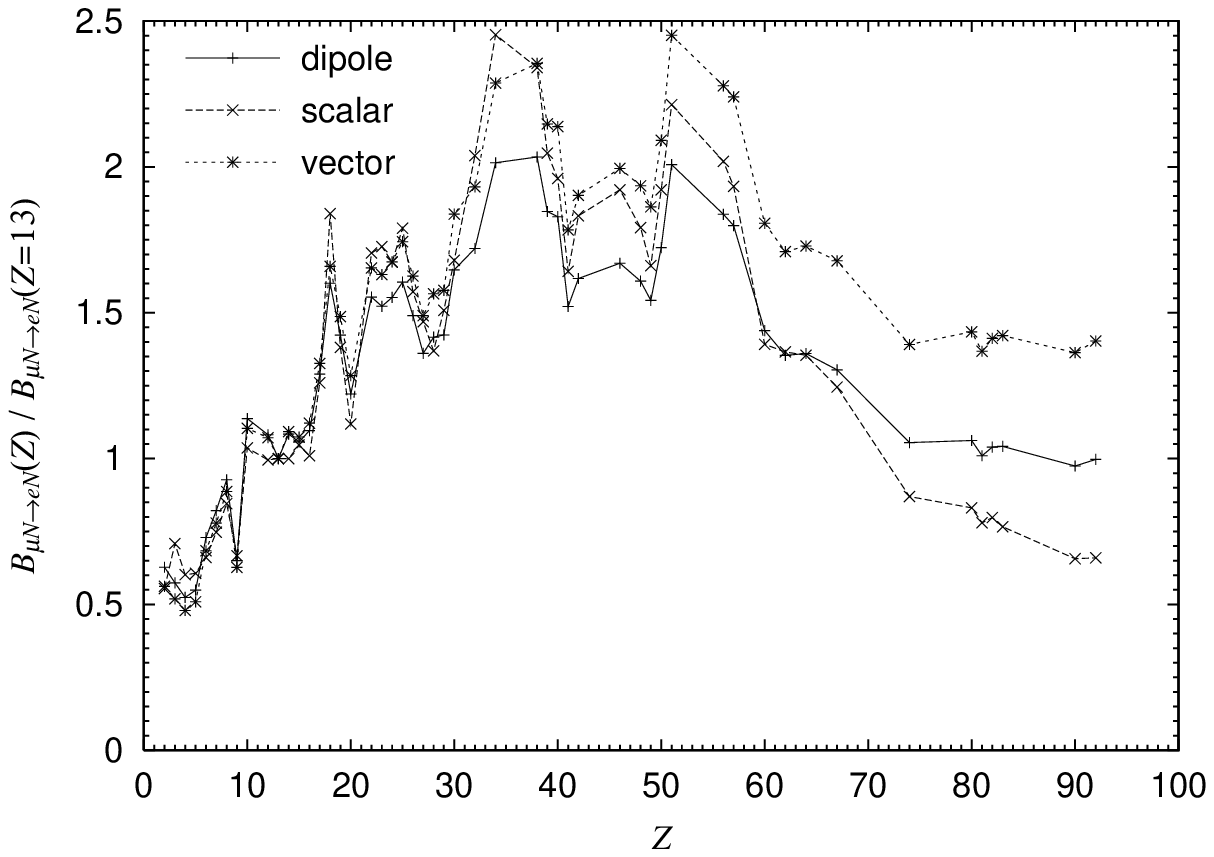} 
    \end{center}
    \caption{
The $\mu$-$e$ conversion ratios for the typical theoretical models are
plotted as functions of the atomic number $Z$.  The solid, the long
dashed, and the dashed lines represent the cases that the photonic
dipole, scalar, and vector operator dominates, respectively.  Proton and
neutron distribution are taken according to method \ref{method:p=n} in
subsection \ref{subsection:nucleon-in-nucleus}, and the conversion
ratios are normalized by the conversion ratio in aluminum nuclei ($Z =
13$).
} \label{fig7}
\end{figure}
\begin{figure}[t]
	\begin{center}
		\includegraphics[width=15cm]{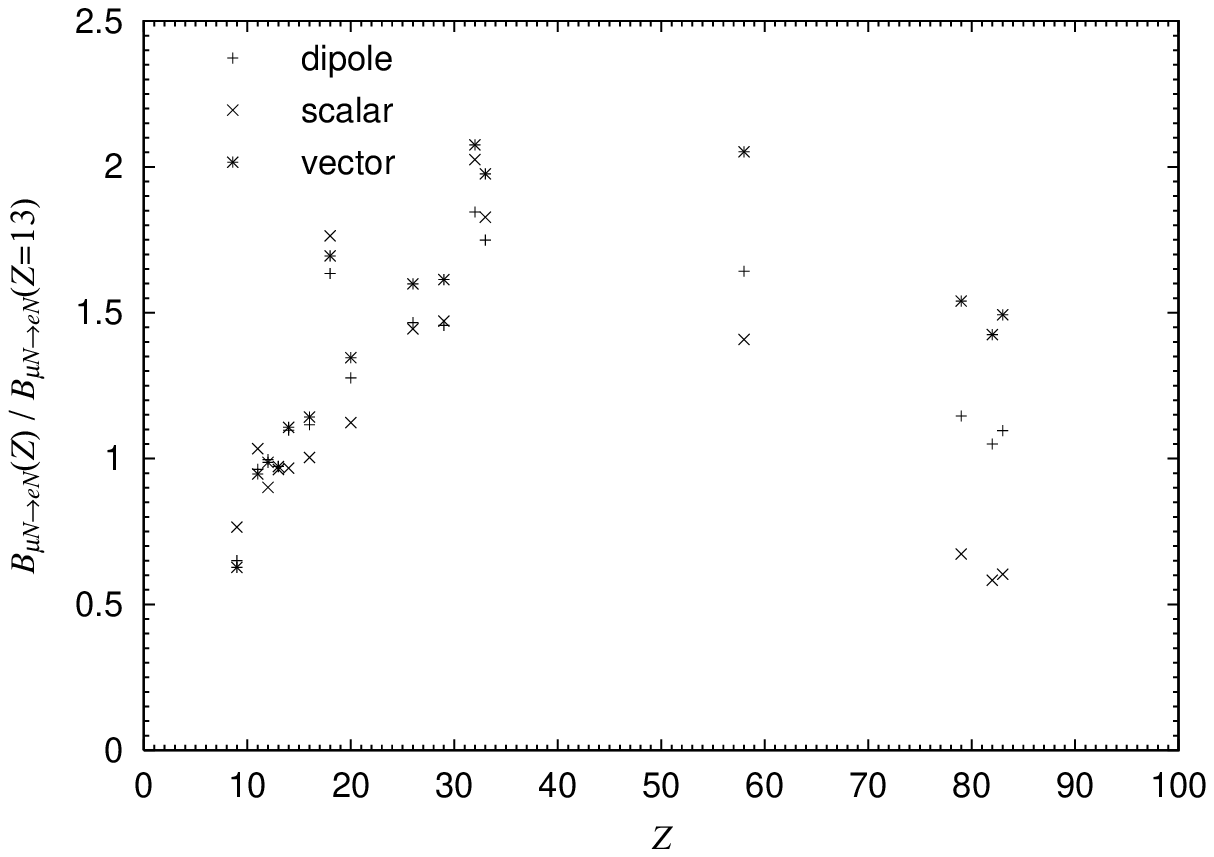} 
	\end{center}
\caption{
The $\mu$-$e$ conversion ratios for the typical theoretical models
evaluated by the method \ref{method:pion} in subsection
\ref{subsection:nucleon-in-nucleus}.  The branching ratio is normalized
by $B_{\mu N \to e N}(Z = 13)$ evaluated by the method \ref{method:p=n}.
} \label{fig:models-pion}
\end{figure}
\begin{figure}[t]
	\begin{center}
		\includegraphics[width=15cm]{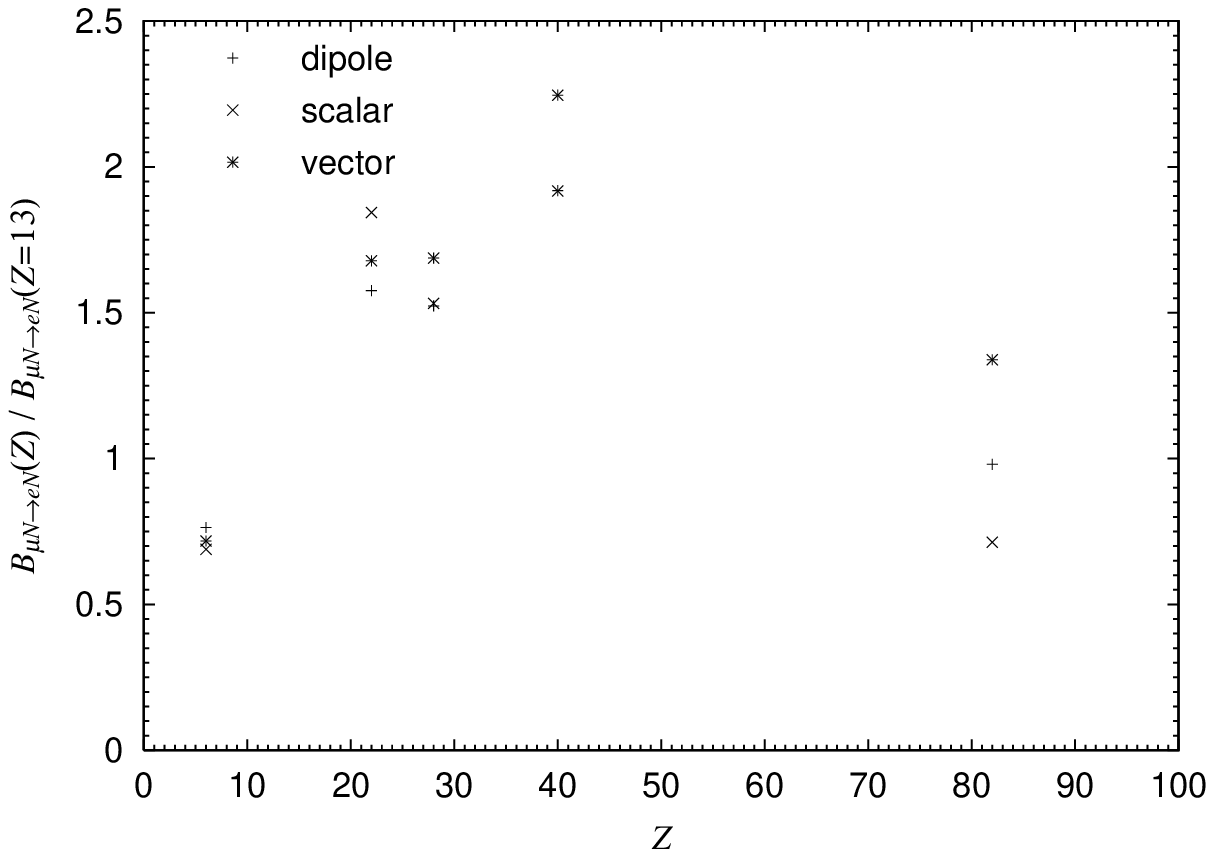} 
	\end{center}
	\caption{ The $\mu$-$e$ conversion ratios for the typical
	theoretical models evaluated by the method \ref{method:proton}
	in subsection \ref{subsection:nucleon-in-nucleus}.  The
	branching ratio is normalized by $B_{\mu N \to e N}(Z = 13)$
	evaluated by the method \ref{method:p=n}.  } 
	\label{fig:models-proton}
\end{figure}

It is now straightforward to evaluate the conversion rate through
Eq.(\ref{eq:omega-final}) once a theoretical model with LFV is
specified.
In this subsection, we present a $\mu$-$e$ conversion branching ratio
for various types of LFV interactions in order to show the possibility
to discriminate the different models through the $Z$ dependence.
We also compare our results with the existing calculations for the 
case where the photonic dipole operators are non-vanishing.

We consider the following three cases:
\begin{enumerate}
    \item
	\label{case:photonic}
	The photonic dipole operators $A_L$ and/or $A_R$ are
	non-vanishing.  The $\mu$-$e$ conversion branching ratio is
	given by
	\begin{eqnarray}
		B_{\mu N \to e N}
		\equiv 
		\frac{\omega_{\rm conv}}{\omega_{\rm capt}}
		=
		\frac{ 2 G_{\rm F}^{2} D^{2} ( |A_{L}|^{2} + |A_{R}|^{2} )}
		{\omega_{\rm capt}}
		\ ,
	\end{eqnarray}
	where $\omega_{\rm capt}$ is the muon capture rate.  For
	convenience, we list the capture rate in Appendix B
	\cite{Suzuki:1987jf}.
    \item
	\label{case:scalar}
    The scalar operators $g_{RS(d)}$ and/or $g_{LS(d)}$ are 
    non-vanishing.
	The $\mu$-$e$ conversion branching ratio in this case is given by
	\begin{equation}
		B_{\mu N \to e N}
		=
		\frac{
		2 G_{\rm F}^{2} 
		( G_{S}^{(d, p)} S^{(p)} + G_{S}^{(d, n)} S^{(n)} )^{2}
		( |g_{LS(d)}|^{2} + |g_{RS(d)}|^{2} )
		}{\omega_{\rm capt}}.
		\label{eq:branch-scalar}
	\end{equation}
    \item
	\label{case:vector}
	The vector operators $g_{RV(u)}$ and $g_{LV(u)}$ satisfies
	$g_{RV(u)} = -2 g_{RV(d)} \neq 0$ and/or $g_{LV(u)} = -2 g_{LV(d)}
	\neq 0$ \ .  The $\mu$-$e$ conversion branching ratio in this case
	is given by
	\begin{equation}
		B_{\mu N \to e N}
		=
		\frac{
		2 G_{\rm F}^{2} V^{(p)2}
		( |\tilde g_{LV}^{(p)}|^{2} + |\tilde g_{RV}^{(p)}|^{2} )
		}{\omega_{\rm capt}}.
		\label{eq:branch-vector}
	\end{equation}
\end{enumerate}
The first case appears as a good approximation in SUSY models for many
cases, especially in SO(10) SUSY GUT models \cite{Barbieri:1995tw} and
in SUSY models with right-handed neutrinos 
\cite{models-Right-Handed-nu}.
The second case is realized in some cases of SUSY models with R-parity
violation \cite{models-R-parity}.
The third case corresponds to the situation where the monopole form
factors give dominant contributions in the $\mu$-$e$-$\gamma$
transition.
The $\mu$-$e$ conversion branching ratio are shown for three cases in
Figs.\ref{fig7} (method \ref{method:p=n}), \ref{fig:models-pion}
(method \ref{method:pion}), and \ref{fig:models-proton} (method
\ref{method:proton}).  In these figures the branching ratios are
normalized by the value for aluminum evaluated by the method
\ref{method:p=n}, namely 
\begin{eqnarray}
    & \mbox{Dipole:} & B_{\mu N \rightarrow e N}  (Z = 13) =
    9.9 \, \left( |A_{L}|^{2} + |A_{R}|^{2} \right),
    \label{eq:numeral-dipole}  \\
    & \mbox{Scalar:} & B_{\mu N \rightarrow e N}  (Z = 13) =
    1.7 \times 10^{2} \, \left( |g_{LS(d)}|^{2} + |g_{RS(d)}|^{2} \right),
    \label{eq:numeral-scalar}  \\
    & \mbox{Vector:} & B_{\mu N \rightarrow e N}  (Z = 13) =
    2.0 \, \left( |\tilde g_{LV}^{(p)}|^{2} + |\tilde g_{RV}^{(p)}|^{2} 
    \right).
    \label{eq:numeral-vector}
\end{eqnarray}
We can see that, for all three types, the branching ratio increases as
$Z$ for $Z \lesssim 30$, are largest for $30 \lesssim Z \lesssim 60$,
and decreases for $Z \gtrsim 60$.
It is also seen that the conversion ratios have large differences in
heavy nuclei depending on the three types of interaction.
From this property we may be able to distinguish models beyond the SM
through several experiments with different targets.

\begin{figure}[t]
    \begin{center}
	\includegraphics[width=15cm]{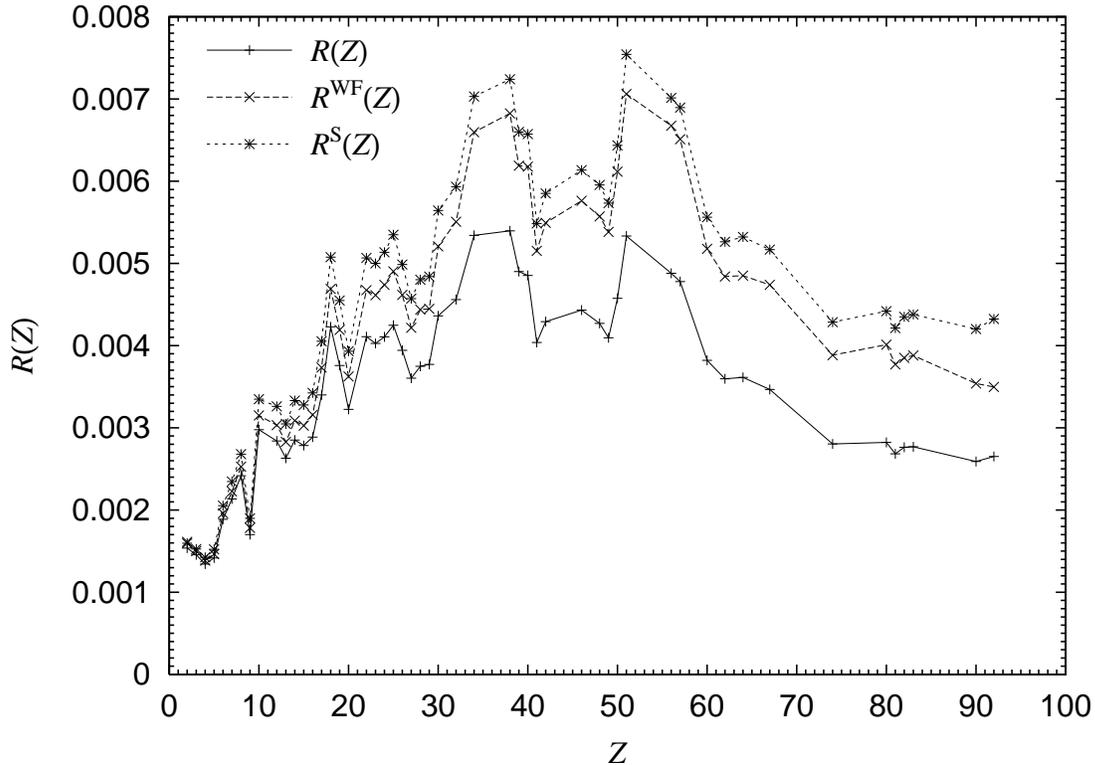} 
    \end{center}
    \caption{%
	The $\mu$-$e$ conversion branching ratio divided by the $\mu \to e
	\gamma$ decay branching ratio for method \ref{method:p=n} is
	plotted as a function of atomic number $Z$.  The solid line
	($R(Z)$), the long-dashed line ($R^{\rm WF} (Z)$), and the dashed
	line ($R^{\rm S} (Z)$) represent the results of our calculation,
	the Weinberg-Feinberg formula, and the Shanker's approximation,
    respectively.%
    }
    \label{fig6}
\end{figure}

In order to see improvements of the present method from older
calculations, we compare three different approximations for the case
where the photonic dipole operators are non-vanishing: namely, our
calculation, Weinberg-Feinberg approximation, and the approximation by
Shanker.  For this purpose we define the ratio $R(Z) \equiv B_{\mu N
\to e N} / B(\mu \to e \gamma)$,
where the $\mu \to e \gamma$ branching ratio $B(\mu \to e \gamma)$ is
given by $384 \pi^2 ( |A_L|^2 + |A_R|^2)$.  The present method thereby
gives
\begin{equation}
	R(Z) = \frac{G_{\rm F}^2 D^2}{192 \pi^2 \omega_{\rm capt}}\ .
\end{equation}
In the Weinberg-Feinberg calculation, the relativistic effects and the
Coulomb distortion were ignored \cite{WF}.
We define the conversion branching ratio $B^{\rm WF}_{\mu N \to e N}$
and the ratio of ratios $R^{\rm WF}(Z)$ in the Weinberg-Feinberg
approximation by the following formula:
\begin{eqnarray}
    B^{\rm WF}_{\mu N \to e N}
    & = &
    \frac{8 G_{\rm F}^2 m_\mu^5}{\pi^2}
    \alpha^{3} Z_{\rm eff}^{4} Z F_{p}^{2} ( |A_L|^2 + |A_R|^2 )
    \frac{1}{\omega_{\rm capt}},
    \label{eq:B_{WF}} \\
    R^{\rm WF}(Z)
    & = &
    \frac{B^{\rm WF}_{\mu N \to e N}}{B(\mu \to e \gamma)}
    =
    \frac{G_{\rm F}^2 m_{\mu}^{5} \alpha^{3} Z_{\rm eff}^{4} Z F_{p}^{2}}
    {48 \pi^{4} \omega_{\rm capt}}
    \ .
    \label{eq:R^{WF}}
\end{eqnarray}
Notice that these are not exactly the same as the formula given in the
original paper because they used approximate formulae for the capture
rate and the form factors for the general photonic transition.
Shanker improved the Weinberg-Feinberg formula taking into account the
relativistic effects and the Coulomb distortion.
In his approximation, the branching ratio and the ratio of ratios for
the dipole photonic interaction are given by
\begin{eqnarray}
    B^{\rm S}_{\mu N \to e N}
    & = &
    128 e^{2}
	G_{\rm F}^2 V^{(p) 2} ( |A_L|^2 + |A_R|^2 ) 
    \frac{1}{\omega_{\rm capt}},
    \label{eq:B^{S}} \\
    R^{\rm S}(Z)
    & = &
    \frac{B^{\rm S}_{\mu N \to e N}}{B(\mu \to e \gamma)}
    =
	\frac{ e^{2} G_{\rm F}^2 V^{(p)2} }{ 3 \pi^{2} \omega_{\rm capt} }
    \ .
    \label{eq:R^{S}}
\end{eqnarray}
We present our $R(Z)$, $R^{\rm WF} (Z)$, and $R^{\rm S} (Z)$ in
Fig.\ref{fig6}.  Here we used the proton density in the method
\ref{method:p=n} and the muon capture rate $\omega_{\rm capt}$ from
the experiments \cite{Suzuki:1987jf}.
We see that the three quantities have similar $Z$ dependence: they
range from 0.002 to 0.006, and are largest for $Z = 30$ -- $60$.  The
values of $R^{\rm WF} (Z)$ and $R^{\rm S} (Z)$ are larger than our
$R(Z)$ by 30\% for $Z \gtrsim 50$.  We have reproduced with a good
accuracy the result by Czarnecki {\it et al.}, where they evaluated
$R(Z)$ for aluminum (Al), titanium (Ti), and lead (Pb) nuclei.
Kosmas obtained in Ref.\cite{Kosmas:2001ij} the result that $R(Z)$ is
a monotonically increasing function, but he did not take into account
the Coulomb distortion effect.  We could indeed obtain the increasing
function of $R(Z)$ by ignoring this effect.  Thus the Coulomb
distortion effect of the wave function is important in the calculation
of the conversion rate for heavy nuclei, as noted by Shanker
\cite{Shanker:1979ap}.

\section{Summary}
\label{section:summary}

We have calculated the coherent $\mu$-$e$ conversion rate for general
LFV interactions for various nuclei.  We have used updated nuclear
data for proton and neutron distributions and taken into account the
ambiguity associated with neutron distribution.  We gave the list of
the overlap integrals in Tables \ref{tab1}, \ref{table:Overlap-pion},
and \ref{table:Overlap-proton} for various nuclei, from which we can
calculate conversion rates for general interactions for LFV using
Eq.(\ref{eq:omega-final}).
We also investigate the $Z$ dependence of the conversion rate.  We see
that the branching ratio increases for the light nuclei such as $Z
\lesssim 30$, are largest for $Z = 30$ -- $60$, and decreases for
heavy nuclei with $Z \gtrsim 60$.  Although this tendency of $Z$
dependence is the same for different types of coupling constants,
there are significant differences in $Z$ dependence of branching
ratios.
We show that the ambiguity in the calculation of the overlap integrals
associated with proton densities ($D$, $S^{(p)}$, and $V^{(p)}$) is
quite small because the charge densities of nuclei are well known.
On the other hand, the overlapping integrals $S^{(n)}$ and $V^{(n)}$
contain uncertainly from the neutron distribution, especially for
heavy nuclei.
We have estimated $S^{(n)}$ and $V^{(n)}$ with several inputs.
Using the neutron density distribution determined from proton
scattering experiments performed in 1970's and pionic atom
experiments, we showed that the conversion rate changes by 20\% --
30\% for heavy nuclei.  Adopting the improved neutron density
distribution determined by the new proton scattering experiment, we
found that the ambiguity is significantly reduced down to a few
percent.
Because the main ambiguity for heavy nuclei is associated with the
neutron density, it will be possible to make a precise prediction if
we can determine the neutron density with improved analysis and
experiments.

The results of our calculation are useful to choose the appropriate
target nuclei for future experiments for the $\mu$-$e$ conversion 
search.
In addition, it may be possible to identify the theoretical models
beyond the Standard Model through the $Z$ dependence of different
interactions when the signal of the $\mu$-$e$ conversion is
experimentally observed.


\section*{Acknowledgments}

We would like to thank J.~Hisano for useful discussions.
This work was supported by the JSPS Research Fellowships
for Young Scientists (R.~K. and M.~K.).
The work of Y.~O. was supported in part by a Grant-in-Aid of the 
Ministry of Education, Culture, Sports, Science and Technology, 
Government of Japan (No.~13640309), priority area ``Supersymmetry and 
Unified Theory of Elementary Particles'' (No.~707).


\appendix
\section{Proton and neutron densities in nuclei}
\label{appendix:nucleon-densities}

We introduce models of nucleon densities in nuclei
and list the values of parameters of these models
used in the calculation.

We used one of the following models for each nucleus \cite{table}.
\begin{enumerate}
    \item
    Harmonic oscillator model (HO):
    \begin{eqnarray}
	\rho_{p(n)} (r)
	=
	\rho_{0} \left[ 1 + \alpha \left( \frac{r}{a} \right)^{2} \right]
	\exp \left[ - \left( \frac{r}{a} \right)^{2} \right]
	\ .
    \end{eqnarray}
    \item
    Two-parameter Fermi model (2pF):
    \begin{eqnarray}
	\rho_{p(n)} (r) =
	\frac{\rho_0}{1+ \exp[ (r-c_{p(n)})/ z_{p(n)}]}
	\ .
	\label{eq:2pF-def}
    \end{eqnarray}
    \item
    Three-parameter Fermi model (3pF):
    \begin{eqnarray}
	\rho_{p(n)} (r)
	=
	\frac{\rho_0 (1 + w_{p(n)} r^2 / c^2_{p(n)})}
	{1 + \exp[ (r - c_{p(n)}) / z_{p(n)}]}\ .
    \end{eqnarray}
    \item
    Two-parameter Gaussian model (2pG):
    \begin{equation}
	\rho_{p(n)} (r)
	=
	\frac{\rho_0}{1 + \exp[ (r^2 - c^2_{p(n)}) / z^2_{p(n)}]}\ .
    \end{equation}
    \item
    Three-parameter Gaussian model (3pG):
    \begin{eqnarray}
	\rho_{p(n)} (r)
	=
	\frac{\rho_0 (1 +w_{p(n)} r^2 / c^2_{p(n)})}
	{1 + \exp[ (r^2-c^2_{p(n)})/ z^2_{p(n)}]}\ .
    \end{eqnarray}
\end{enumerate}
Here $c_{p(n)}$, $z_{p(n)}$, and $w_{p(n)}$
are the model parameters and $\rho_0$ is the normalization
constant.
We also used the following model-independent analysis for several
nuclei.
\begin{enumerate}
    \addtocounter{enumi}{5}
    \item
    The Fourier-Bessel expansion (FB):
    \begin{eqnarray}
	\rho_{p(n)} (r) = \left \{
	\begin{array}{lll}
	    \displaystyle
	    \sum_v a_v j_0(v \pi r / R)& {\rm for} & r \leq R\\
	    0& {\rm for} & r > R\\
	\end{array}
	\right.
	\ ,
    \end{eqnarray}
	where $a_v$ are the coefficients, $R$ is the cutoff radius, and
	the function $j_0 (z) = \sin z / z$ is the spherical Bessel
	function of the zeroth order.
    \item
    The Sum of Gaussian expansion (SOG):
    \begin{equation}
	\rho_{p(n)} (r)
	=
	\sum_{i}
	A_{i}
	\left\{
	\exp \left[
	- \left( \frac{r - R_{i}}{\gamma} \right)^{2}
	\right]
	+
	\exp \left[
	- \left( \frac{r + R_{i}}{\gamma} \right)^{2}
	\right]
	\right\},
	\label{eq:SOG-def}
    \end{equation}
    where
    \begin{equation}
	A_{i} =
	\frac{Z e Q_{i}}{2 \pi^{3/2} \gamma^{3} (1 + 2 R_{i}^{2} / 
	\gamma^{2})}.
	\label{eq:SOG-norm-def}
    \end{equation}
\end{enumerate}

We list the model and its parameters used in calculation in Table
\ref{tab_nucl}.  We do not list parameters for FB and SOG there; see 
Ref.\cite{table}.

\begin{table}[p]
    \begin{center}
	\begin{tabular}{|c||c|l|l|l|}
	    \hline
	    Nucleus &
	    Model & $c_p$ or $a_{p}$ (fm) & $z_p$ (fm) or $\alpha$
	    & $w_p$
	    \\ 
	    \hline \hline
	    $^{4}_{2}$He & SOG & --- & --- & --- \\
$^{7}_{3}$Li & HO & 1.77 & 0.327 & --- \\
$^{9}_{4}$Be & HO & 1.791 & 0.611 & --- \\
$^{11}_{5}$B & HO & 1.69 & 0.811 & --- \\
$^{12}_{6}$C & FB & --- & --- & --- \\
$^{14}_{7}$N & 3pF & 2.570 & 0.5052 & -0.180 \\
$^{16}_{8}$O & FB & --- & --- & --- \\
$^{19}_{9}$F & 2pF & 2.59 & 0.564 & --- \\
$^{20}_{10}$Ne & 3pF & 2.791 & 0.698 & -0.168 \\
$^{24}_{12}$Mg & 3pF & 2.791 & 0.698 & -0.168 \\
$^{27}_{13}$Al & FB & --- & --- & --- \\
$^{28}_{14}$Si & 3pF & 3.340 & 0.580 & -0.233 \\
$^{31}_{15}$P & FB & --- & --- & --- \\
$^{32}_{16}$S & FB & --- & --- & --- \\
$^{35}_{17}$Cl & 3pF & 3.476 & 0.599 & -0.10 \\
$^{40}_{18}$Ar & FB & --- & --- & --- \\
$^{39}_{19}$K & 3pF & 3.743 & 0.585 & -0.201 \\
$^{40}_{20}$Ca & FB & --- & --- & --- \\
$^{48}_{22}$Ti & FB & --- & --- & --- \\
$^{51}_{23}$V & 2pF & 3.91 & 0.532 & --- \\
$^{52}_{24}$Cr & FB & --- & --- & --- \\
$^{55}_{25}$Mn & 2pF & 3.89 & 0.567 & --- \\
$^{56}_{26}$Fe & 3pG & 3.505 & 2.325 & 0.380 \\
$^{59}_{27}$Co & 2pF & 4.158 & 0.575 & --- \\
$^{58}_{28}$Ni & FB & --- & --- & --- \\
$^{63}_{29}$Cu & 2pF & 4.218 & 0.596 & --- \\
$^{64}_{30}$Zn & 3pG & 3.664 & 2.425 & 0.342 \\
$^{74}_{32}$Ge & FB & --- & --- & --- \\
$^{80}_{34}$Se & 2pG & 4.622 & 2.365 & --- \\
$^{88}_{38}$Sr & FB & --- & --- & --- \\
$^{89}_{39}$Y & 2pF & 4.86 & 0.542 & --- \\
$^{90}_{40}$Zr & FB & --- & --- & --- \\

	    \hline 
	\end{tabular}
    \end{center}
    \caption{
	The model parameters of the proton density functions are
	listed. These values are extracted from Refs.\cite{table}.  The
	abbreviations HO, 2pF, 3pF, 2pG, 3pG, FB, and SOG represent the
	harmonic oscillator model, the two-parameter Fermi model, the
	three-parameter Fermi model, the two-parameter Gaussian model,
	the three-parameter Gaussian model, the Fourier-Bessel
	expansion, and sum of Gaussian, respectively.  We do not list
	here parameters for FB and SOG; see Ref.\cite{table}.
}
	\label{tab_nucl}
\end{table}
\addtocounter{table}{-1}
\begin{table}[p]
    \begin{center}
	\begin{tabular}{|c||c|l|l|l|}
	    \hline
	    Nucleus &
	    Model & $c_p$ or $a_{p}$ (fm) & $z_p$ or $\alpha$ (fm) 
	    & $w_p$
	    \\ 
	    \hline \hline
	    $^{93}_{41}$Nb & 2pF & 4.953 & 0.541 & --- \\
$^{98}_{42}$Mo & FB & --- & --- & --- \\
$^{110}_{46}$Pd & FB & --- & --- & --- \\
$^{114}_{48}$Cd & 2pF & 5.314 & 0.571 & --- \\
$^{115}_{49}$In & 2pF & 5.357 & 0.563 & --- \\
$^{120}_{50}$Sn & 3pG & 5.110 & 2.619 & 0.292 \\
$^{121}_{51}$Sb & 2pF & 5.32 & 0.57 & --- \\
$^{138}_{56}$Ba & 3pG & 5.3376 & 2.6776 & 0.3749 \\
$^{139}_{57}$La & 2pF & 5.71 & 0.535 & --- \\
$^{142}_{60}$Nd & 2pF & 5.839 & 0.569 & --- \\
$^{152}_{62}$Sm & FB & --- & --- & --- \\
$^{158}_{64}$Gd & FB & --- & --- & --- \\
$^{165}_{67}$Ho & 2pF & 6.12 & 0.57 & --- \\
$^{166}_{68}$Er & 3pF & 5.98 & 0.446 & 0.19 \\
$^{181}_{73}$Ta & 2pF & 6.38 & 0.64 & --- \\
$^{184}_{74}$W & 2pF & 6.51 & 0.535 & --- \\
$^{197}_{79}$Au & 2pF & 6.38 & 0.535 & --- \\
$^{204}_{80}$Hg & FB & --- & --- & --- \\
$^{205}_{81}$Tl & SOG & --- & --- & --- \\
$^{208}_{82}$Pb & FB & --- & --- & --- \\
$^{209}_{83}$Bi & FB & --- & --- & --- \\
$^{232}_{90}$Th & 2pF & 6.851 & 0.518 & --- \\
$^{238}_{92}$U & 2pF & 6.874 & 0.556 & --- \\

	    \hline 
	\end{tabular}
    \end{center}
    \caption{%
    (Continued).
    }
    \label{tab_nucl2}
\end{table}

\section{Muon capture rate in nuclei}
\label{appendix:capture-rate}

We list in Table \ref{tabcapt} the muon capture rates $\omega_{\rm
capt}$ which are used in our calculation \cite{Suzuki:1987jf}.

\begin{table}[p]
    \begin{center}
	\begin{tabular}{|c|l||c|l|}
	    \hline
	    Nucleus &
	    $\omega_{\rm capt}$ ($10^6 {\rm s}^{-1}$) &
	    Nucleus &
	    $\omega_{\rm capt}$ ($10^6 {\rm s}^{-1}$) \\
	    \hline \hline
	    $^{4}_{2}$He & 0.000336 & $^{72}_{32}$Ge & 5.569 \\
$^{7}_{3}$Li & 0.0018 & $^{75}_{33}$As & 6.104 \\
$^{9}_{4}$Be & 0.0074 &  $^{88}_{38}$Sr & 7.02 \\
$^{11}_{5}$B & 0.0219 &  $^{89}_{39}$Y & 7.89 \\
$^{12}_{6}$C & 0.0388 &  $^{90}_{40}$Zr & 8.66 \\
$^{14}_{7}$N & 0.0693 &  $^{93}_{41}$Nb & 10.36 \\
$^{16}_{8}$O & 0.1026 &  $^{92}_{42}$Mo & 9.614 \\
$^{19}_{9}$F & 0.229 &  $^{110}_{46}$Pd & 10.00 \\
$^{20}_{10}$Ne & 0.200 &  $^{114}_{48}$Cd & 10.61 \\
$^{23}_{11}$Na & 0.3772 &  $^{115}_{49}$In & 11.40 \\
$^{24}_{12}$Mg & 0.4841 &  $^{120}_{50}$Sn & 10.44 \\
$^{27}_{13}$Al & 0.7054 &  $^{121}_{51}$Sb & 10.21 \\
$^{28}_{14}$Si & 0.8712 &  $^{138}_{56}$Ba & 9.94 \\
$^{31}_{15}$P & 1.1185 &  $^{139}_{57}$La & 10.71 \\
$^{32}_{16}$S & 1.352 &  $^{140}_{58}$Ce & 11.60 \\
$^{35}_{17}$Cl & 1.333 &  $^{142}_{60}$Nd & 12.50 \\
$^{40}_{18}$Ar & 1.30 &  $^{152}_{62}$Sm & 12.22 \\
$^{39}_{19}$K & 1.849 &  $^{158}_{64}$Gd & 11.82 \\
$^{40}_{20}$Ca & 2.557 &  $^{165}_{67}$Ho & 12.95 \\
$^{48}_{22}$Ti & 2.59 &  $^{166}_{68}$Er & 13.04 \\
$^{51}_{23}$V & 3.069 &  $^{181}_{73}$Ta & 12.86 \\
$^{52}_{24}$Cr & 3.472 &  $^{184}_{74}$W & 12.36 \\
$^{55}_{25}$Mn & 3.857 &  $^{197}_{79}$Au & 13.07 \\
$^{56}_{26}$Fe & 4.411 &  $^{205}_{81}$Tl & 13.90 \\
$^{59}_{27}$Co & 4.940 &  $^{207}_{82}$Pb & 13.45 \\
$^{58}_{28}$Ni & 5.932 &  $^{209}_{83}$Bi & 13.10 \\
$^{63}_{29}$Cu & 5.676 &  $^{232}_{90}$Th & 13.1 \\
$^{64}_{30}$Zn & 5.834 &  $^{238}_{92}$U & 12.4 \\

	    \hline 
	\end{tabular}
    \end{center}
    \caption{
    The total capture rates used in calculation are listed
    \cite{Suzuki:1987jf}.
    }
    \label{tabcapt}
\end{table}

\end{document}